%% file: main.tex
\documentclass[twocolumn,twocolappendix,usenames,dvipsnames]{aastex63}
\usepackage{capt-of}

\input{citation_fix}

\newcommand{\mach}{\mathcal{M}}
\newcommand{\macha}{\mathcal{M}_\mathrm{A}}

\usepackage{verbatim}
\usepackage{amsmath}
\usepackage{multirow}
\usepackage{relsize}
\usepackage{graphicx}
\usepackage{xcolor}
\usepackage{appendix}

\setcitestyle{aysep={}} 
\shorttitle{The Density PDF and the Gas Compression and Expansion Rates}
\shortauthors{Appel et al.}

\graphicspath{{./}{figures/}}

\begin{document}

\title{What Sets the Star Formation Rate of Molecular Clouds? The Density Distribution as a Fingerprint of Compression and Expansion Rates}

\author[0000-0002-6593-3800]{Sabrina M. Appel}
\email{appel@physics.rutgers.edu}
\affiliation{Department of Physics and Astronomy, 
Rutgers University,
136 Frelinghuysen Rd., 
Piscataway, NJ 08854, USA}

\author[0000-0001-5817-5944]{Blakesley Burkhart}
\affiliation{Department of Physics and Astronomy, 
Rutgers University,
136 Frelinghuysen Rd., 
Piscataway, NJ 08854, USA}
\affiliation{Center for Computational Astrophysics, 
Flatiron Institute, 
162 Fifth Avenue, 
New York, NY 10010, USA}

\author[0000-0002-6648-7136]{Vadim A. Semenov}
\affiliation{Center for Astrophysics $|$ Harvard \& Smithsonian, 
60 Garden St., 
Cambridge, MA 02138, USA}

\author[0000-0002-0706-2306]{Christoph Federrath}
\affiliation{Research School of Astronomy and Astrophysics, 
The Australian National University, 
Canberra, ACT 2611, 
Australia}
\affiliation{Australian Research Council Centre of Excellence in All Sky Astrophysics (ASTRO3D), Canberra, ACT 2611, Australia}

\author[0000-0003-4423-0660]{Anna L. Rosen}
\affiliation{Center for Astronomy \& Space Sciences, University of California, San Diego, La Jolla, CA 92093, USA}

\author[0000-0002-3389-9142]{Jonathan C. Tan}
\affiliation{Department of Space Earth \& Environment, Chalmers University of Technology, Gothenburg, SE-412 96, Sweden}
\affiliation{Department of Astronomy, University of Virginia, Charlottesville, VA 22904, USA}

\begin{abstract}
We use a suite of 3D simulations of star-forming molecular clouds, with and without stellar feedback, magnetic fields, and driven turbulence, to study the compression and expansion rates of the gas as functions of density. 
We show that, around the mean density, supersonic turbulence promotes rough equilibrium between the amounts of compressing and expanding gas, consistent with continuous gas cycling between high and low density states. 
We find that the inclusion of protostellar jets produces rapidly expanding and compressing low-density gas.
We find that the gas mass flux peaks at the transition between the lognormal and power-law forms of the density probability distribution function (PDF).
This is consistent with the transition density tracking the post-shock density, which promotes an enhancement of mass at this density (i.e., shock compression and filament formation). 
At high densities, the gas dynamics are dominated by self-gravity: the compression rate in all of our runs matches the rate of the run with only gravity, suggesting that processes other than self-gravity have little effect at these densities.
The net gas mass flux becomes constant at a density below the sink formation threshold, where it equals the star formation rate. 
The density at which the net gas mass flux equals the star formation rate is one order of magnitude lower than our sink threshold density, corresponds to the formation of the second power-law tail in the density PDF,
and sets the overall star formation rates of these simulations. 

\end{abstract}


\section{Introduction} \label{sec:intro}

Stars form from the gravitational collapse of cold, dense gas within giant molecular clouds (GMCs), which are supersonically turbulent, magnetized, and self-gravitating \citep[e.g.,][]{Padoan1997, Mckee_Ostriker2007, KE2012, Myers14a, KrumholzBurkhart2018}. 
Models of star formation must include how self-gravity and magnetohydrodynamical processes interact with stellar feedback and the galactic environment to produce gas that collapses into stars \citep{Collins12a, Padoan12a, Burkhart2017}. 
To this end, the volume density probability distribution function (PDF) is a commonly used tool for analytic models of star formation that provides insight into the distribution of gas within GMCs \cite[e.g.,][]{KrumholzMcKee2005, Padoan+2011, Hennebelle+2008, Hennebelle09a, HennebellChabrier2011, Padoan+2011, hopkins12, FederrathKlessen2012, Padoan14a, Burkhart2018}. 
Models based on the density PDF have been useful in understanding a wide variety of processes, and are also used in subgrid models for galaxy simulations \citep[e.g.,][]{Braun+2015,Semenov+2016,Semenov+2021,Trebitsch2017,Rosdahl2018,Kretschmer2020,Gensior+2020}.  

The shape of the density PDF for gas within GMCs (for both volume densities and column densities) has been well studied \citep[see e.g.,][]{vazquezsemadeni97, Passot+2003, Klessen2000, Federrath2008, Kritsuk+2011, Collins12a, Burkhart12,FederrathKlessen2013, Imara2016,Chen2018,Appel+2022,Ma+2022}. 
It is well understood that supersonic turbulence produces a lognormal distribution in the density PDF and that self-gravity produces a power-law tail in dense, star-forming gas \citep[see e.g.,][]{Burkhart2018, BurkhartMocz2019, JaupartChabrier2020}.
At the highest densities, the density and column density PDFs show evidence of an additional power-law tail  \citep{Schneider+2015,Khullar2021}. 
Simulations suggest that the first power-law forms due to collapse under the influence of self-gravity while the second forms at higher densities due the formation of accretion disks around newly forming stars \citep{Khullar2021}.

Previous work has also shown how the shape of the density PDF changes with the inclusion of different physical processes including supersonic turbulence, self-gravity, magnetic fields, and stellar feedback, as well as how the PDF changes in time as gas is exchanged between low- and high-density regions and eventually converted to stars \citep[e.g.,][]{Burkhart2018, BurkhartMocz2019, Appel+2022}. 
In particular, \cite{Appel+2022} showed that the inclusion of protostellar outflows can produce an excess of diffuse gas in the volume density PDF.
That paper also showed how diffuse, turbulently supported gas (the lognormal portion of the density PDF) is transferred into dense, gravitationally collapsing gas (the power-law portion of the density PDF), which in turn collapses into stars.

The density PDF alone gives a limited indication as to the dynamics of the gas. 
\cite{Appel+2022} used an analysis of the time evolution of the density PDF to provide insight regarding how gas flows between different density regimes, which traced the dynamical evolution of the gas. 
In particular, \cite{Appel+2022} showed that the high-density power-law tail of the density PDF is stable in time, suggesting that the diffuse gas rapidly replenishes the power-law tail as the dense gas is converted into or accreted by protostars. 
This suggests a connection between the density PDF and the gas dynamics that we explore further in this paper.

In this work, we explore the gas dynamics in a suite of 3D magnetohydrodynamic (MHD) simulations of star-forming regions; we connect the density PDF of simulated star-forming regions to the compression and expansion rates of the gas, and we compare these rates to the star formation rates of the simulated regions.
In particular, we quantify the gas dynamics using the divergence of the gas velocities ($\nabla \cdot \vec{v}$, where $\vec{v}$ is the gas velocity), which traces the expanding (positive divergence) and gravitationally collapsing gas (negative divergence), respectively.
Hence, the velocity divergence can be used as a metric to understand the compression and expansion rates of the gas, and can be used to quantify how the gas dynamics affect the star formation rate of star-forming molecular clouds.
As we will show in this work, the compression and expansion rates can be converted to a net gas mass flux from low to high densities, which can be directly compared to the star formation rate of the simulation. 
By examining both the compression and expansion rates and the gas mass flux as functions of density, we are able to compare these quantities to the shape of the density PDF to provide  insight into the impacts of various physical processes including the influence of magnetic fields, turbulence, and stellar feedback, on the gas dynamics.

This paper is organized as follows. Section~\ref{sec:simulations} describes the simulations that are used in our analysis.
Section~\ref{sec:analysis} reviews the theory of the density PDF and how it connects to the star formation rates (SFR), and introduces our model for calculating the compression and expansion rates that we use in our analysis. 
Section~\ref{sec:results} describes our simulation results and our analysis of the gas dynamics in our simulations (i.e., the compression and expansion rates and gas mass flux).
Finally, we discuss our results in Section~\ref{sec:discussion}, and our key conclusions are summarized in Section~\ref{sec:conclusion}.

\section{Simulations and Numerical Parameters}\label{sec:simulations} 

We use a suite of 5 simulations that were run using the FLASH MHD code \citep{Fryxell2000}, where each simulation increases in complexity by including additional physical processes -- gravity, turbulence driving, magnetic fields (i.e., MHD), and stellar feedback -- so that we can separate how each of these processes affects the SFR, the gas density distribution, and the gas dynamics. 
The first 4~simulations in this suite are identical in their basic parameters to the study in \citet{Federrath2015}, which compared the SFR between runs with 1) gravity only, 2) as 1, but also including turbulence, 3) as 2, but also including magnetic fields, and 4) as 3, but also including jet/outflow feedback. 
Here we redo these runs (with different turbulence seeds and at higher resolution), and we add a 5th run, which is set up the same as run~4, but also includes protostellar heating feedback. 
Details of the runs are provided below.

\begin{deluxetable}{lc}[b]
\tabletypesize{\footnotesize}
\tablecaption{Overview of key simulation parameters
\label{tab:sims_const}}
\tablecolumns{5}
\tablewidth{0pt}
\tablehead{
\colhead{Parameter} &
\colhead{Value}
}
\startdata
Mean density ($\rho_0$)  &  $3.28 \times 10^{-21}$ g cm $^{-3}$ \\
Initial total mass ($M_{\mathrm{tot}}$) & 388 M$_{\odot}$ \\
Box size ($L$) & 2 pc \\
Velocity dispersion ($\sigma_{v}$) & 1 $\mathrm{km \, s}^{-1}$ \\
Sonic Mach number ($\mach_s$) & 5 \\
Driving parameter ($\zeta$) & 0.5  \\
Virial parameter ($\alpha_{\mathrm{vir}}$) & 1.0 \\
Magnetic field ($B$) & 10 $\mu$G \\
Alfv\'{e}nic Mach number ($\macha$) & 2.0 \\
Max.~effective resolution ($N_{\mathrm{res}}^{3}$) & $1024^3$
 \enddata
\tablecomments{Note that quantities such as the mean density, total mass, and magnetic field strength refer to the initial values for each run and the turbulence parameters (velocity dispersion, sonic Mach number, and driving parameter) refer to the values that result from the turbulence driving method.
}
\end{deluxetable}

\subsection{Simulation Code and Initial Conditions}

FLASH is a publicly available hydrodynamics code that uses adaptive mesh refinement (AMR) \citep{BergerColella1989} and can include many interoperable modules \citep{Fryxell2000, Dubey+2008}.
The MHD solver used to run the simulations presented here uses a Godunov-type method using a second-order, five-wave approximate Riemann solver, termed HLL5R \citep{Waagan+2011}. The self-gravity of the gas is modelled with a multi-grid Poisson solver \citep{Ricker2008}.

Table~\ref{tab:sims_const} summarizes the key parameters of our simulation suite.
Each simulation has a computational box size of 2~pc and periodic boundary conditions.
The initial total gas mass for each simulation is 388~M$_{\odot}$ and each run is initialized with a uniform density of $\rho_0 = 3.28 \times 10^{-21}$~g~cm$^{-3}$. This corresponds to a virial parameter of $\alpha_{\mathrm{vir}} = 1.0$ after turbulence driving has established a fully-developed turbulent cloud with a velocity dispersion of 1~km/s (details on the driving method are provided below).

\subsection{Grid Refinement}

Two levels of AMR are used. The base grid (level~1) has a uniform resolution of $512^3$ grid cells to capture the turbulence well \citep{Kitsionas+2009,Federrath+2010,PriceFederrath2010,Kritsuk+2011codes}. 
At the highest level of AMR (level~2), the cell size is $402.9$~AU, corresponding to a maximum effective grid resolution of $N_{\mathrm{res}}^{3} = 1024^3$ cells. 
Refinement is based on the Jeans length, requiring that the simulations resolve the Jeans length by at least 30~grid cells at any time and any point in space \citep{Federrath+2011}. 
This is a much higher refinement of the Jeans length than required to simply avoid artificial fragmentation \citep{Truelove97a}, which only requires 4~cells per Jeans length. 
The reason to prefer a Jeans resolution of at least 30~grid cells is that it allows us to capture solenoidal motions and minimum magnetic field amplification via the turbulent dynamo process \citep{Federrath2016b} on the Jeans scale \citep{Federrath+2011}.

\begin{figure*}[t!]
\centering
\includegraphics[width = \linewidth]{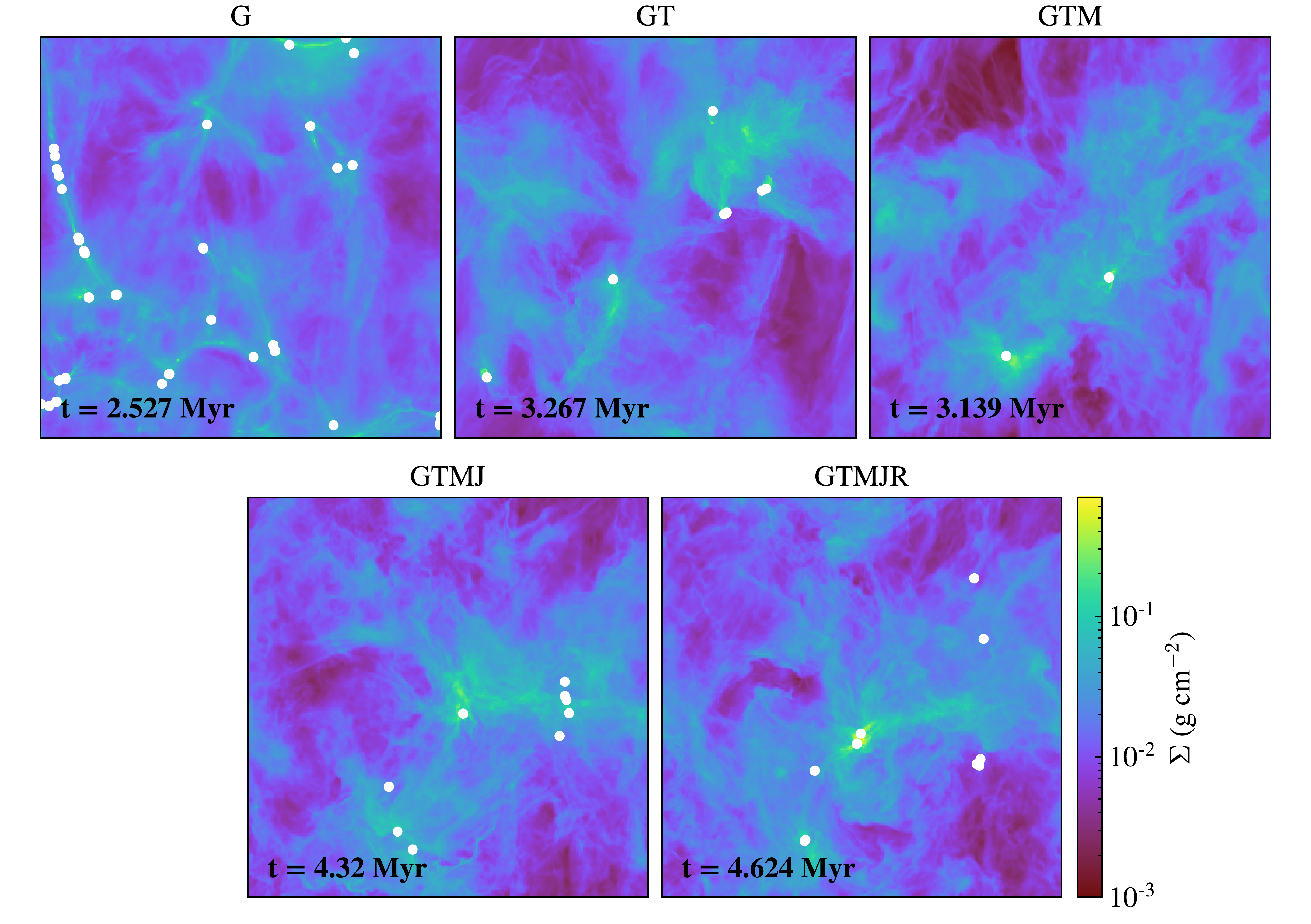}
\caption{Density projection plots for each of the simulations described in Table~\ref{tab:sims} at snapshots where the SFE is approximately 5\%. The corresponding times are shown on the plot. The projected positions of sink particles are shown as white circles.
\label{fig:projections}}
\end{figure*}

\begin{deluxetable*}{llccccccc}[b!t]
\centering
\tabletypesize{\footnotesize}
\captionof{table}{Summary of simulations
\label{tab:sims}}
\tablecolumns{6}
\tablewidth{0pt}
\tablehead{
\colhead{} &
\colhead{Simulation} &
\colhead{Turbulence?} &
\colhead{B-fields?} &
\colhead{Jets?} &
\colhead{Heating Feedback?} &
\colhead{SFR (M$_{\odot}$ yr$^{-1}$)} &
\colhead{$N_{\mathrm{sinks}}$} &
\colhead{$\rho_{*}$ (g cm$^{-3}$)}
}
\startdata
1 & \textsc{G}  & no & no & no & no  & $12.7 \times 10^{-5}$   & 75 & N/A \\
2 & \textsc{GT} & yes & no & no & no & $5.60 \times 10^{-5}$   & 20 & $8.03\times 10^{-19}$  \\
3 & \textsc{GTM}& yes & yes & no & no & $2.58 \times 10^{-5}$  & 4  & $5.38\times 10^{-19}$ \\
4 & \textsc{GTMJ} & yes & yes & yes & no & $1.50 \times 10^{-5}$   & 18 & $1.20 \times 10^{-18}$ \\
5 & \textsc{GTMJR} & yes & yes & yes & yes & $1.40 \times 10^{-5}$  & 22 & $1.79 \times 10^{-18}$
 \enddata
\tablecomments{All simulations include self-gravity. Turbulence indicates whether turbulence continues to be driven throughout the simulation run. When turbulence driving is turned off, the velocity field is also set to zero, as discussed below.
B-fields indicates whether magnetic fields are included.
Jets refers to the inclusion of protostellar outflows, and Heating Feedback refers to the inclusion of radiative feedback in the form of radiative heating \citep{MathewFederrath2020}. 
SFR refers to the average star formation rate for each simulation between SFE~$=2\%$ and SFE~$=10\%$ (the SFE range used in Figs.~\ref{fig:medians-net} through~\ref{fig:net-flux}) and is the value plotted in Figs.~\ref{fig:median-sfr}~and~\ref{fig:net-flux}. N$_{\mathrm{sinks}}$ indicates the final number of sinks each simulation formed by SFE $=10\%$. The $\rho_{*} = \rho_{0} \, e^{s_{*}}$ values are the densities at which the net gas mass flux matches the SFR, as discussed in Section~\ref{sec:sfr_comparison}, and are shown as black stars in Fig.~\ref{fig:net-flux}.}
\end{deluxetable*}

\subsection{Star Formation}

To model the formation of stars in each simulation, we use the sink particle method described in \cite{Federrath+2010}. 
When gas collapses under the influence of gravity and the density in a given cell reaches the sink formation density threshold, a sink particle is formed (if the gas also meets several additional criteria, as discussed below). 
Sink particles represent a star-disk system and only form on the highest AMR level.
While the sink particles represent unresolved star-disk systems, we are unable to discern how much mass is in each component and, therefore, when we calculate the SFR (in Section~\ref{sec:results}), we assume the sink mass is representative of the star.

The sink formation density is \citep{Federrath+2010,Federrath+2014}:
\begin{equation}
    \rho_{\mathrm{sink}} = \frac{\pi c_{\mathrm{s}}^{2} }{G \lambda_{\mathrm{J}}^{2}} = \frac{\pi c_{\mathrm{s}}^{2} }{4 \, G r_{\mathrm{sink}}^{2}} 
    \label{eq:sinkdens}
\end{equation}
where $c_\mathrm{s}$ is the sound speed, $G$ is the gravitational constant, $\lambda_{\mathrm{J}}$ is the local Jeans length, and $r_{\mathrm{sink}}$ is the sink radius. 
The sink particle radius is set to 2.5 grid cell lengths \citep{Federrath+2010,Federrath+2014} to avoid artificial fragmentation \citep{Truelove97a}. 
Before a sink particle is formed from gas above the density threshold given by Eq.~\ref{eq:sinkdens}, the gas surrounding the cell that exceeds this density undergoes a series of checks for collapse. 
Only if the gas is bound and collapsing within a Jeans length centred on the cell that exceeds $\rho_{\mathrm{sink}}$, is a sink particle created. 
This procedure avoids artificial creation of sink particles in unbound regions where the density can exceed $\rho_{\mathrm{sink}}$ merely due to shock compression rather than gravitational collapse \citep[see quantifications and discussions in][]{Federrath+2010}.

Existing sink particles can accrete gas if that gas is within $r_{\mathrm{sink}}$ of the sink particle, above $\rho_{\mathrm{sink}}$, and bound and collapsing toward the sink particle. 
The gravitational interactions between gas and sink particles and between the sink particles themselves is computed by direct summation using spline gravitational softening within $r_{\mathrm{sink}}$. 
Please see \citet{Federrath+2010,Federrath+2014} for details on the sink particle method.

\subsection{Turbulence Driving}

To drive turbulence we use the methods described in \citep{Federrath+2010}, with the code publicly available on GitHub \citep{Federrath+2022_turb}. 
In summary, this uses an Ornstein-Uhlenbeck (OU) process to evolve the Fourier modes of the turbulent acceleration field used to drive the turbulence in real space. 
We generate Fourier modes in an interval between wave numbers $k/(2\pi/L)=1$ and $3$, with amplitudes following a parabola, such that the peak of the parabola is at $k/(2\pi/L)=2$ and the driving amplitude is exactly zero at $k/(2\pi/L)=1$ and $3$. 
Based on this, the driving scale is at $k/(2\pi/L)=2$ in Fourier space, i.e., at $\ell_\mathrm{turb}=L/2=1\,\mathrm{pc}$ in real space. 
The amplitude is adjusted such that the resulting velocity dispersion is $v_\mathrm{turb}=1\,\mathrm{km/s}$ (which corresponds to a sonic Mach number of $\mathcal{M}=5$, considering the sound speed $c_\mathrm{s}=0.2~\mathrm{km/s}$ for molecular gas at about 10\,K) in the fully-developed turbulent state.

This defines a turbulent turnover time, as the combination of the driving scale and the target velocity dispersion, namely $t_\mathrm{turb} = \ell_\mathrm{turb} / v_\mathrm{turb} \sim 1\,\mathrm{Myr}$. 
The auto-correlation timescale of the OU process is set to this timescale, producing a turbulent acceleration field that varies smoothly in space and time. 
The turbulence is fully developed (resulting in a fully-developed turbulent cloud) after about $2 \, t_\mathrm{turb}$ \citep{Federrath+2010,PriceFederrath2010}, at which point we activate gravity and allow for star formation based on the criteria defined in the previous subsection. 
This procedure is identical to previous works \citep[e.g.,][]{FederrathKlessen2012,Federrath2015,Nam+2021,MathewFederrath2021}.

Here we use a natural mixture of solenoidal and compressive modes in the turbulence driving (acceleration) field \citep{Federrath2008}. 
This is controlled by performing a Helmholtz decomposition of the acceleration field in Fourier space, and mixing the modes together, as desired \citep[see details in][which defines the turbulence driving mode mixture parameter $\zeta$]{Federrath+2010}. 
A purely solenoidal driving field is obtained with $\zeta=1$ and a purely compressive driving field is obtained with $\zeta=0$. 
The natural mixture of driving modes corresponds to $\zeta=0.5$, not to be confused with the ratio of the density dispersion to sonic Mach number, referred to as $b=\sigma_\rho/(\rho_0 \, \mathcal{M})$ and where $b\sim0.38$ in the case of naturally mixed driving \citep[see Fig.~8 in][]{Federrath+2010}, which is an indirect consequence of the driving on the density and velocity fields, while $\zeta$ is the parameter that directly controls the mode mixture in the turbulence driving (acceleration) field.

All simulations discussed here use exactly the same turbulence driving sequence and random seed to drive the turbulence. 
However, we note that when magnetic fields are included, the density structure at the start of the star-formation stage of the simulations is not identical to the density field without magnetic fields, as the fields naturally alter the density and velocity structure of the clouds. 
Apart from that, however, runs~1~(\textsc{G}) and~2~(\textsc{GT}) (both without magnetic fields), and runs~3~(\textsc{GTM}), 4~(\textsc{GTMJ}), and 5~(\textsc{GTMJR}) (all with magnetic fields), respectively, have exactly identical density and velocity fields at the beginning of the star-formation stage (with the exception of run~1 (\textsc{G}), where the velocity field is completely removed at the start of the star-formation stage). 
Each of these runs and the differences in their setups are discussed below and summarized in Tab.~\ref{tab:sims}

\subsection{Model Variations}

Fig.~\ref{fig:projections} shows density projection plots for all 5 of our simulations, and Table~\ref{tab:sims} summarizes the parameters that change between each simulation, as well as selected key results, such as the average SFR.

The first simulation (\textsc{G}) only includes self-gravity.
To set up this simulation run, turbulence is driven as described above for two turbulent turnover times to fully establish the turbulence, at which point turbulence driving is stopped, all gas velocities are set to 0, and the cloud is set to be in pressure equilibrium (to have constant pressure throughout the simulation domain while keeping $\gamma=1.1$). 
Finally, self-gravity is turned on.
This setup ensures an identical initial density distribution between run \textsc{G} and the subsequent run \textsc{GT} (which includes turbulent velocities and driving), before allowing the gas to collapse under the influence of only gravity, without any contribution from either turbulent driving or decaying turbulence.

The second simulation (run \textsc{GT}) is initialized identically to run \textsc{G}. However, unlike run \textsc{G}, we do not alter the gas velocities or gas pressure after the initial driving period, and the turbulent driving continues with the same turbulent driving parameter and sonic Mach number once gravity is turned on.

The third simulation (run \textsc{GTM}) is initialized identically to run \textsc{GT}, except for the addition of a magnetic field.
Following the setup described in \citet{Federrath2015} and \citet{MathewFederrath2021}, the initial magnetic field is assumed to be uniform along the $\hat{z}$ direction where $B_{z} = 10\,\mu\mathrm{G}$. 
The turbulent driving during the initial two turnover times mixes the magnetic field orientation such that the field is no longer uniform by the time gravity is turned on, and an additional turbulent magnetic field component is established \citep{Federrath2016b}. 
The total magnetic field, together with the turbulent velocity field, corresponds to an Alfv\'en Mach number of 2 in the fully-developed turbulent regime.

The final two simulations are identical to run \textsc{GTM}, but also include stellar feedback. 
In runs \textsc{GTMJ} and \textsc{GTMJR}, the sink particles produce two-component protostellar outflows that consist of a fast collimated jet and a slower, wide-angle outflow similar to observed disk winds. 
This two-component model approximates the overall features of observed jets and is described in detail in \citet{Federrath+2014}. 
Run \textsc{GTMJR} additionally includes radiative heating from the sink particles, as first introduced in \cite{Federrathetal2017}, and updated in \cite{MathewFederrath2020}.
The polar heating model described in \citet{MathewFederrath2020} sets up the radiation to be predominantly bipolar rather than isotropic since the sink particle represents a star-disk system and the disk will shield equatorial radiation \citep{Rosen2016}.
This radiative feedback model only includes heating (and neglects photoionization and radiation pressure). 
However, for low-mass star formation, radiative heating is the dominant effect of radiative feedback, and this model compares reasonably well with low-mass star-formation simulations that include radiative transfer \citep{Offner09a}. 
As we will show, the highest mass sink particle that forms in run \textsc{GTMJR} (the only run that implements this radiative feedback model) is $5.6$\,M$_{\odot}$. 
Therefore, we conclude that only including radiative heating is a reasonably accurate description of radiative feedback for this run \citep{MathewFederrath2020}.

\section{Analysis Methods}  \label{sec:analysis} 

\cite{Appel+2022} investigated both the shape of the density PDF and how different parts of the density distribution evolve in time, beginning to give insight into the dynamics of the gas.
In this work, we continue to investigate the gas dynamics by considering the gas compression and expansion rates as a function of density.  
In this section, we first introduce the density PDF, and then we introduce the calculation of the compression and expansion rates, which we will use to study the gas dynamics that can be related to the SFR.

\subsection{The Density PDF}  \label{sec:pdf} 

Many analytic models of star formation assume a time-independent, lognormal density PDF, as suggested by the statistics of supersonic turbulence \citep[e.g.,][]{KrumholzMcKee2005, Padoan+2011, HennebellChabrier2011, FederrathKlessen2012}.
However, recent work suggests a piecewise lognormal plus power-law density PDF is a better fit for gas that includes both turbulence and gravity \citep[e.g.,][]{Imara2016, Chen2018,Khullar2021,Ma+2022}.
The piecewise density PDF as proposed by \cite{Burkhart2018} takes the form:
\begin{equation}
    p_{\mathrm{LN+PL}}(s) = 
\begin{cases}
 N\frac{1}{\sqrt{2\pi}\sigma_s}e^{-\frac{( s - s_0)^2}{2\sigma_s^2}}  & s < s_{\mathrm{t}} \\
 N C e^{-\alpha s}  & s > s_{\mathrm{t}} \ ,
\end{cases}
\end{equation}
where $\sigma_s$ and $s_0$ are the width and the mean of the lognormal portion, respectively, and $\alpha$ is the slope of the power-law portion.

As discussed in Section~\ref{sec:intro}, the density PDF is a key component of many analytic star formation models, including sub-grid models for the SFR. 
In these models, the density PDF is used to quantify the gas mass fraction that can form stars (because it is dense enough to collapse), and, weighted by a free-fall time factor, predict the SFR. 
The density PDF therefore quantifies how much of the gas is primarily turbulent versus how much gas is collapsing. 

The SFR can be calculated by integrating over the density PDF above a critical density for star formation and multiplying by the appropriate timescales and densities \citep[e.g., see][]{KrumholzMcKee2005, Padoan+2011, HennebellChabrier2011, FederrathKlessen2012, Burkhart2018}.
The critical density ($s_{\mathrm{crit}}$) can be defined in a number of ways; \cite{FederrathKlessen2012} give a thorough overview of several commonly used critical densities \citep[see also][and references therein]{Burkhart2018}. 
Ultimately, each definition of the critical density finds a way to characterize the density at which gravity becomes dynamically important. 
For example, \cite{KrumholzMcKee2005} directly compares the scales at which gravity and turbulence become equal, i.e., the sonic scale \citep{Federrath+2021}. Alternatively, \citet{Burkhart2018} suggests that the critical densities discussed in \citet{FederrathKlessen2012} are effectively traced by the transition density ($s_{\mathrm{t}}$) where the lognormal density PDF changes to a power-law PDF. 
The transition between these regimes serves as evidence within the density PDF of the fact that gravity becomes dynamically dominant \citep{Burkhart2018, BurkhartMocz2019}.

In our analysis, we explore how different physical processes impact the gas dynamics and how these dynamics are reflected in the density PDF. 
Thus, we are particularly interested in investigating the density at which self-gravity dominates over other physical processes and how the gas density PDF relates to the SFR. 
Hence, we will investigate if the transition density, $s_{\mathrm{t}}$, from \citet{Burkhart2018} is the density at which gravity becomes dynamically dominant and sets the stage for star formation.

\subsection{Calculating the Compression Rate} \label{sec:rate-method}

The density PDF describes the density distribution of the gas at a single point in time, but it can be connected to the underlying gas dynamics with the continuity equation. 
We quantify the rate of change of the density using the Lagrangian formulation of the continuity equation, which is given by
\begin{align}
    \frac{D \rho}{D t} + \rho  \left( \nabla \cdot \vec{v} \right) &=0 \ \ ,
    \label{eq:conteq_rho}
\end{align}
where we use $D/Dt \equiv \partial/\partial t + \vec{v} \cdot \nabla$ as a shorthand for the Lagrangian derivative (see Appendix~\ref{sec:app_lagrangian} for derivation).

To compare this expression to the density PDF, which we have calculated in terms of the natural logarithm of the normalized density,
\begin{equation}
s=\ln(\rho/\rho_0) \ \ , 
\end{equation}
we rearrange Eq.~\ref{eq:conteq_rho} to be expressed in terms of $s$ (see Appendix~\ref{sec:app_lagrangian} for details):
\begin{equation}
    \frac{D s}{Dt} \equiv - ( \nabla \cdot \Vec{v} \,) \ \ .
\end{equation}
Thus, we have a connection between the time evolution of the natural logarithm of the normalized density (the $Ds/Dt$ term) and the gas dynamics (as represented by the velocity vector, $\Vec{v}$).

Because $s$ is dimensionless, the quantity $Ds/Dt$ has dimensions of inverse time, i.e., it is the rate of change of the logarithmic density contrast $s$.  
This quantity represents the flow rate of gas into or out of a particular region of the simulation. 
\textit{For the purpose of this study, we refer to positive values of $Ds/Dt$ as compression rates and negative values of $Ds/Dt$ as expansion rates.}
Regions with a positive rate ($Ds/Dt > 0$) have net compressing gas, i.e., converging flows ($\nabla\cdot\Vec{v}<0$), which means that more gas is entering that region than is leaving it. 
Conversely, regions with a negative rate ($Ds/Dt < 0$) have net expanding gas, i.e., diverging flows ($\nabla\cdot\Vec{v}>0$), which means that more gas is leaving that region than is entering it. 

We compute the compression rate, $Ds/Dt$, for every simulation cell using the \verb|yt| package \citep{Turk11a}. 
We calculate the gradient of each component of the velocity field, which we then use to construct the divergence of the velocity field
\begin{equation} 
   \frac{D s}{Dt} =  -(\nabla \cdot \Vec{v}) = - \left( \partial_x v_x + \partial_y v_y + \partial_z v_z  \right) \  .
   \label{eq:dsdt}
\end{equation}
Thus, for every cell in the simulation, we now have both a density value and a compression or expansion rate, and we can investigate how the compression and expansion rates behave as functions of time and how they compare to the density PDF.

\section{Results}  \label{sec:results}

\subsection{Star Formation throughout Each Run} \label{sec:overview_sims}  

\begin{figure}[htb!]
\centering
\includegraphics[width = \linewidth]{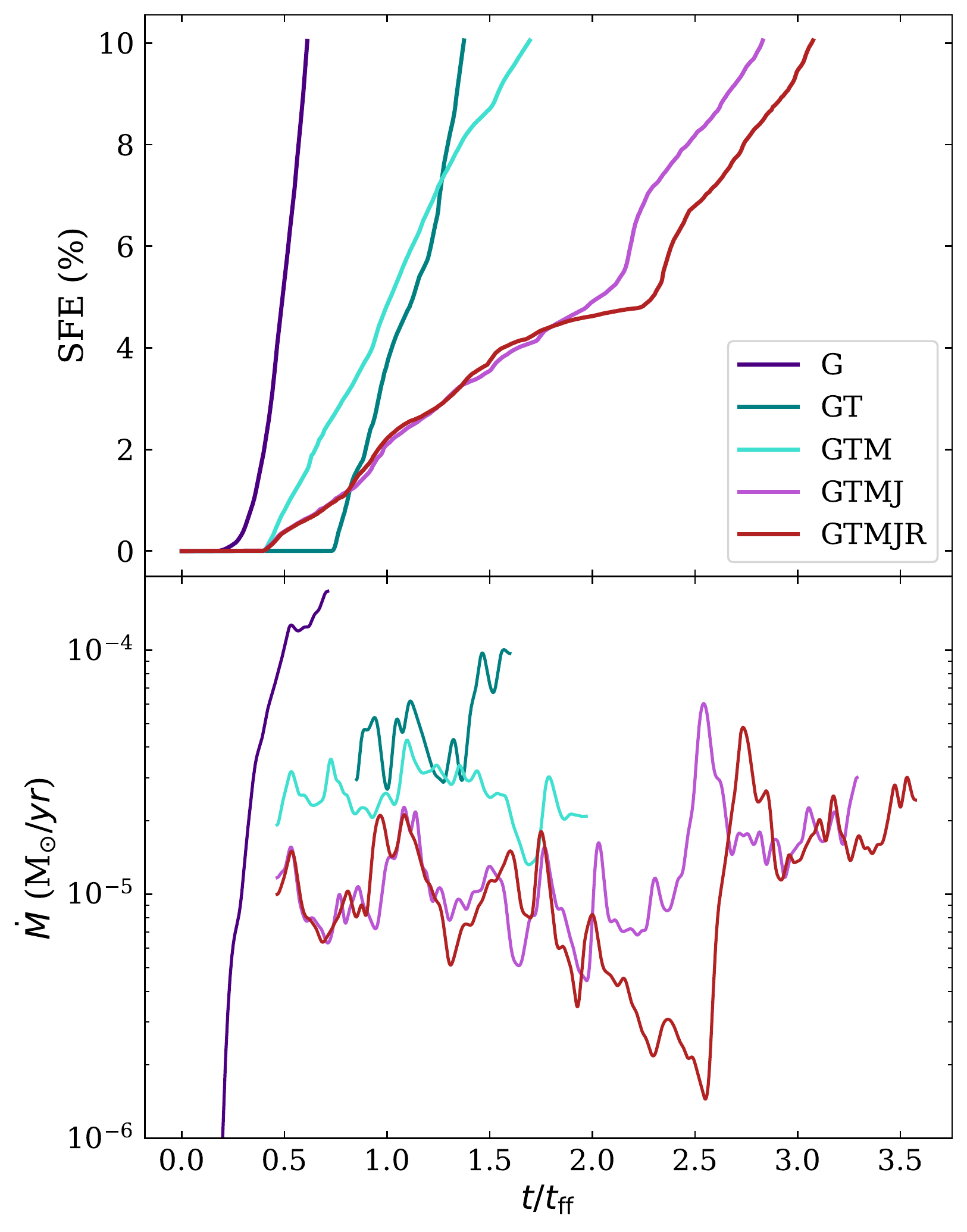}
\caption{The integrated star formation efficiency (SFE; top panel) and the star formation rate (SFR; bottom panel) as a function of time for each of the simulations described in Table~\ref{tab:sims}. The $t_{0} = 0$ point is defined to be the point at which gravity is turned on for each simulation. As additional physical processes are included, the SFR drops and it takes longer for each simulation to reach a similar integrated SFE.
\label{fig:sfe_time}}
\end{figure}

\begin{figure*}[htb!]
\centering
\includegraphics[width = \linewidth]{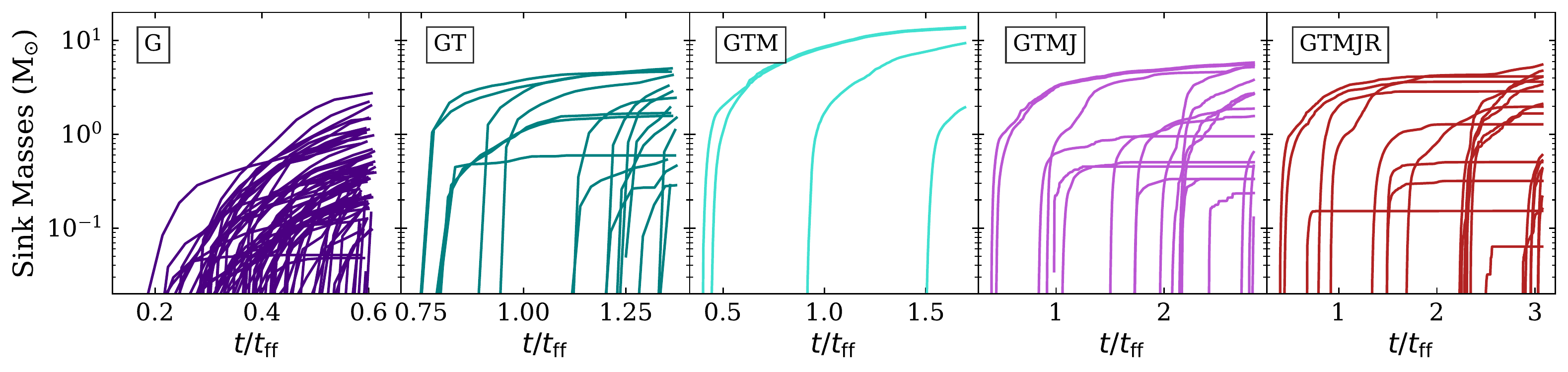}
\caption{The evolution of the masses of individual sink particles as a function of time for each of the simulations described in Table~\ref{tab:sims}. Each sink particle follows a single line on the plot. The $t_{0} = 0$ point is defined to be the point at which gravity is turned on for each simulation. Each run takes a different amount of time to form the first sink particle. Thus, note that the x-axis range varies for each panel.
\label{fig:sinks_time}}
\end{figure*}

We first compare the impact of the different physical processes in each simulation by examining the integrated SFE and the SFR for each simulation.
In the upper panel of Fig.~\ref{fig:sfe_time}, we show the SFE as a function of time for each of the simulations, where the integrated SFE is defined as
\begin{equation}
    \mathrm{SFE} = \frac{M_{*}}{M_{\mathrm{init}}},
\end{equation}
where $M_{*}$ is the total stellar mass formed and $M_{\mathrm{init}}$ is the initial cloud mass.
Fig.~\ref{fig:sfe_time} demonstrates that the SFE evolves at very different rates for each simulation due to the different physical properties that are included in each run.
To account for this difference when the various runs are compared, we use SFE to characterize the evolutionary stage of the simulation in addition to the simulation time. 
For example, Fig.~\ref{fig:projections} shows density projections of snapshots at $\mathrm{SFE}\approx 5\%$. 

The evolution of the SFR -- which directly relates to the slope of the SFE as a function of time -- is plotted in the lower panel of Fig.~\ref{fig:sfe_time}. 
Although there is a lot of variation over time in the SFR for all of the simulations, there is a clear overall decrease in the SFR as more physical processes are included. 
This is also evident in the mean SFR values shown in Table~\ref{tab:sims}. 
This decrease in the SFR with the inclusion of additional physics has been seen and quantified in previous studies \citep[see e.g.,][]{Federrath2015}.

As an additional diagnostic of the variations between each simulation, we consider the masses of the individual sink particles as a function of time. 
In Fig.~\ref{fig:sinks_time}, we plot the mass growth of each individual sink particle as a function of time and Table~\ref{tab:sims} reports the final number of sink particles ($N_{\mathrm{sinks}}$) for each run. 
We see that run \textsc{G} (the far left panel) forms a large number of sink particles, but only forms relatively low mass ($<4$ M$_{\odot}$) sink particles throughout the simulation.
Run \textsc{GT} forms fewer, higher mass sink particles compared to run \textsc{G}. 
The inclusion of magnetic fields results in yet less fragmentation due to additional magnetic support and run \textsc{GTM} forms only four sink particles, including two that reach up to  $\sim$14 M$_{\odot}$.
The inclusion of jet feedback, however, increases fragmentation \citep{Federrath+2014,Guszejnov+2020,MathewFederrath2021} and runs \textsc{GTMJ} and \textsc{GTMJR} both produce a greater number of sink particles that are all $5.8$ M$_{\odot}$ or less. 

From both Fig.~\ref{fig:sfe_time} and Fig.~\ref{fig:sinks_time} we can see that there is very little difference between runs \textsc{GTMJ} and \textsc{GTMJR}. 
The inclusion of radiative feedback in run \textsc{GTMJR}  slightly slows down star formation relative to run  \textsc{GTMJ}, which only includes protostellar jet feedback. 
Similarly, the inclusion of radiative feedback does not substantially change the amount of fragmentation, as seen in Fig.~\ref{fig:sinks_time} and Table~\ref{tab:sims}. 
Thus, we choose to omit run \textsc{GTMJ} in our subsequent analysis, because it is very similar to run \textsc{GTMJR}. 
We include some further discussion of this run in Appendix~\ref{sec:app_J}.

\subsection{Expansion and Compression Rates as Functions of Density}  \label{sec:rate-dens-plot}

\begin{figure*}[htb!]
\centering
\includegraphics[width = \linewidth]{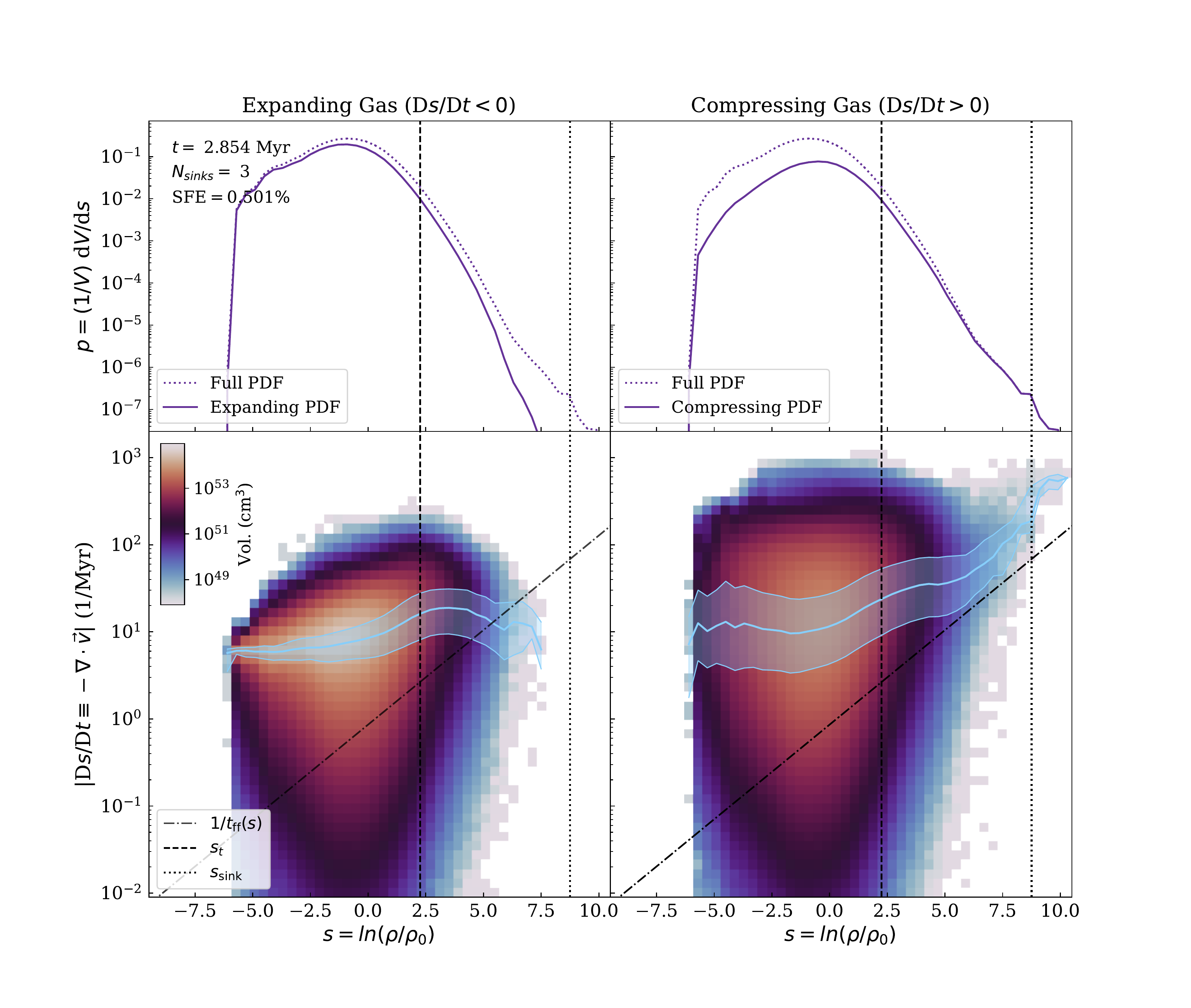} 
\caption{The density PDFs and histograms of the compression and expansion rates for a single snapshot of run \textsc{GT} where the SFE~$\approx 0.5\%$.  \textbf{Top:} The volume-weighted density PDF for all of the gas in the simulation region (dotted line) and the volume-weighted density PDF for only the expanding gas ($Ds/Dt < 0$; left column) or only the compressing gas ($Ds/Dt > 0$; right column) within the simulation region.
\textbf{Bottom:} A 2D, volume-weighted histogram of the expansion rate as a function of density for only the expanding gas ($Ds/Dt < 0$; left column) or of the compression rate as a function of density for only the compressing gas ($Ds/Dt > 0$; right column). The corresponding median rate as a function of density (with the 25th to 75th percentiles) is shown as a blue line (and shaded region). 
The free-fall rate as a function of density is overplotted.
\textbf{All:} For all panels, the transition density ($s_{\mathrm{t}}$) from \cite{Appel+2022} is overplotted as a vertical dashed line.  The sink formation density threshold ($s_{\mathrm{sink}}$) for the simulation is also shown as a dotted vertical line. Key information (time, number of sink particles, and integrated SFE) for the particular snapshot shown here is indicated on the plot.
\label{fig:rate_dens_all}}
\end{figure*}

In Fig.~\ref{fig:rate_dens_all}, we compare the density PDF to the compression and expansion rates for a single snapshot of run \textsc{GT} where the SFE~$\cong 0.5\%$. 
Given the value of $Ds/Dt$ for every cell in the simulation, we can split up all of the gas in the simulation into expanding and compressing gas and consider the density PDF for each component of the gas. 
We plot the compression and expansion rates in separate panels because the collapsing and expanding gas trace different density regimes. 
In particular, the collapsing gas traces higher densities and includes higher rate values than the expanding gas. 
Thus, the top panels of Fig.~\ref{fig:rate_dens_all} shows the separate volume-weighted density PDFs for the expanding gas (solid line; left panel) and  the compressing gas (solid line; right panel).  
For comparison, the overall density PDF (dotted lines) is also shown.
In the bottom panels of Fig.~\ref{fig:rate_dens_all}, we show the compression and expansion rates, $Ds/Dt$, given by Eq.~\ref{eq:dsdt} in units of Myr$^{-1}$ versus $s=\ln (\rho/\rho_0)$ as a 2D histogram (the heat maps in the bottom two panels of Fig.~\ref{fig:rate_dens_all}). 
The left panel shows the 2D histogram of the expansion rate as a function of $s$ of the expanding gas (gas with a negative $Ds/Dt$) and the right panel shows the corresponding histogram for the compression rate of the compressing gas (gas with a positive $Ds/Dt$).
We also show the median compression and expansion rates as a function of density with the interquartile range shown as a shaded region.
For comparison, we show the free-fall rate (i.e., the reciprocal of the gravitational free-fall time) as a function of density, as given by
\begin{equation}
    \frac{1}{t_{\mathrm{ff}} } = \left(\frac{3 \pi}{32 \, G \, \rho }\right) ^{-1/2} \ \ .
    \label{eq:r_ff}
\end{equation}

From this plot, we can see that the expanding gas contributes more to the overall PDF at low densities as compared to the compressing gas. 
Conversely, at high densities, the overall density PDF is primarily composed of compressing gas. 
\textit{Indeed, at densities $s\gtrsim5$, the compressing gas PDF and the overall PDF are essentially identical and the expansion rate dramatically drops off, meaning that compression dominates in this portion of the PDF. 
As we will see in the subsequent analysis, this density is the point at which the gas mass flux equals the SFR in our simulations.}
This is also the density range where we see the development of a second power-law tail in the PDF, in agreement with previous work \citep[][]{Khullar2021}.

In Fig.~\ref{fig:rate_dens_all}, we show the density PDFs for only a single snapshot of a single simulation.  
However, the same trend (e.g., the overall PDF matching the expanding gas PDF at low densities and the compressing gas PDF at high densities) is apparent for all physics cases and throughout the run of the simulations. 
Appendix~\ref{sec:app_PDFs} shows examples of the density PDF for several physics cases and multiple points in time.

Comparison of top and bottom panels shows that some aspects of the distribution of the heat map correspond to the shape of the density PDF.
In particular, there are many fewer counts and less spread in the compressing gas at high densities where the density PDF has a lower value. 
Indeed, there is very little gas at all at high densities in the expanding gas heat map. 
Meanwhile, there is a higher concentration of gas at average to low densities in the expanding gas heat map than in the compressing gas heat map. 
The heat maps emphasize that for both expanding and compressing gas there is a lot of spread in the rate values, with gas at a wide variety of rates at most densities.
However, the median line and the interquartile range indicate that the majority of the gas is clustered around a particular rate for a given density. 
For example, the median expansion rate drops at high densities, while the median compression rate increases at high densities.

\subsection{The Effect of Different Physics on Gas Compression and Expansion Rates} \label{sec:phys_comparison}

\begin{figure*}[htb!]
\centering
\includegraphics[width = 0.75\linewidth]{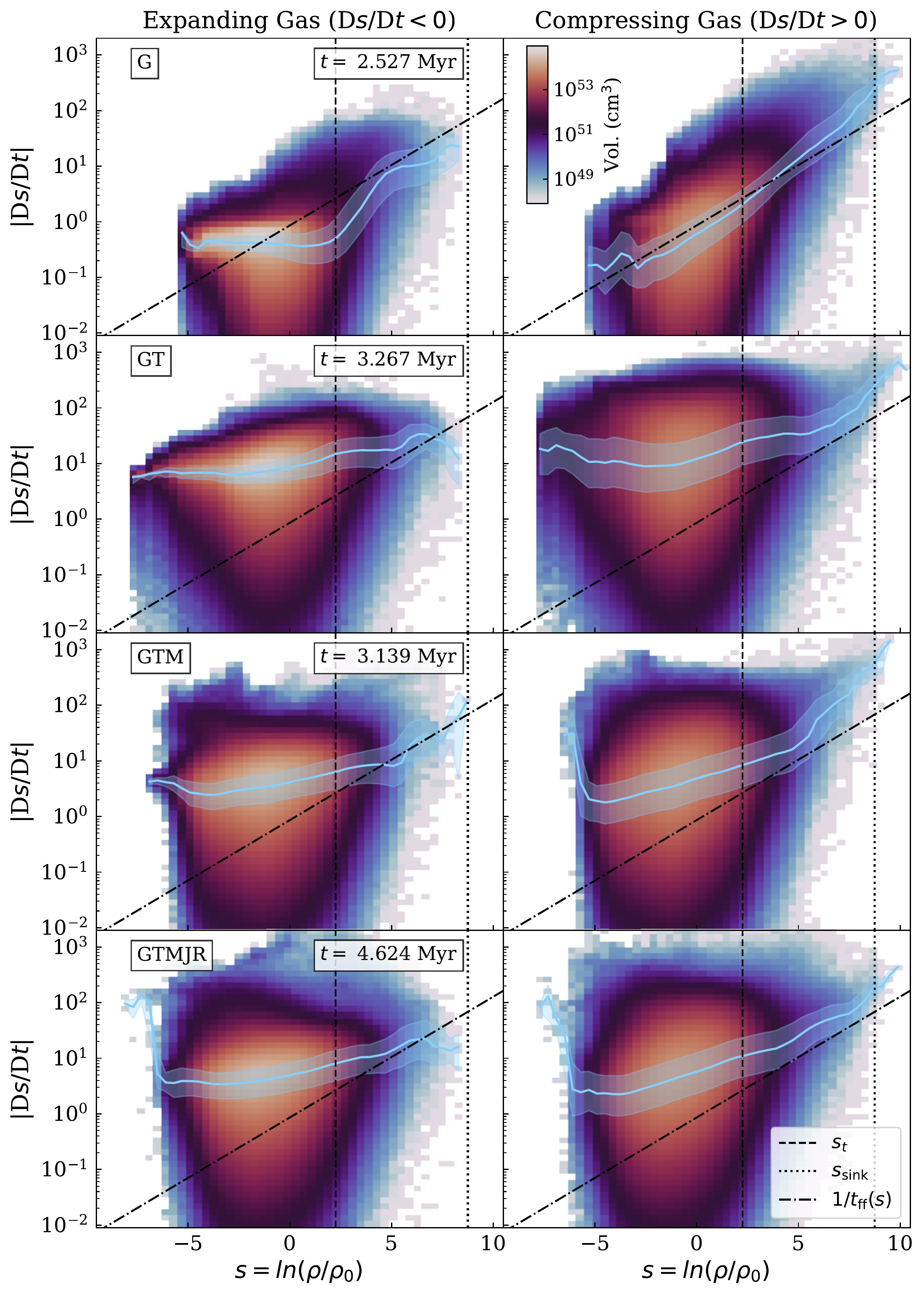}
\caption{Each row is the same as the bottom panels of Fig.~\ref{fig:rate_dens_all} and shows the 2D, volume-weighted histograms of the expansion and compression rates as a function of density. Each row shows a single snapshot of a different simulation. The snapshots chosen for this plot all have SFE~$\approx 5\%$ and the time since gravity was turned on is shown on the plot for each snapshot.
\label{fig:rate_dens_sum}}
\end{figure*}

Figure~\ref{fig:rate_dens_sum} shows the rate-density plots for our four main simulations (runs \textsc{G}, \textsc{GT}, \textsc{GTM}, and \textsc{GTMJR}).
All four physics cases are shown at an approximate midpoint of each simulation, corresponding to SFE$\cong5\%$ (the corresponding times are shown on the plot). 
Again, we show the expanding gas for a given snapshot in the left-hand panels and the compressing gas in the right-hand panels.

Progressing from the top of the figure to the bottom, we can see how the rate-density plot changes as different physical processes are added.
The top row of Fig.~\ref{fig:rate_dens_sum} shows run \textsc{G}, which only includes self-gravity. 
The rate-density plot for this case shows a few distinctive features. 
First, the median line of the compressing gas follows or is slightly above the free-fall rate at all densities. 
In contrast, the expanding gas has a relatively flat distribution around the mean density and only increases toward the free-fall rate at higher densities.

However, in the second row of Fig.~\ref{fig:rate_dens_sum}, we see that the median rate for run \textsc{GT} has an elevated and approximately flat distribution for both the expanding and compressing gas at all densities below the transition density (vertical dashed line), suggesting that turbulence reduces the density dependence of the compression and expansion rates below $s_{\mathrm{t}}$.
Above $s_{\mathrm{t}}$, the compressing gas rises faster than the free-fall rate while the expanding gas remains flat before turning slightly down.
This suggests that for gas below $s_{\mathrm{t}}$, turbulence increases both the rate of convergence and the rate of divergence of the gas relative to the free-fall rate. 
Above $s_{\mathrm{t}}$, the behavior of the compressing gas is mostly determined by gravity.

In the third row of Fig.~\ref{fig:rate_dens_sum}, we see that the inclusion of magnetic fields has a less prominent effect than the difference between run \textsc{G} and run \textsc{GT}. 
However, it still produces some differences in the rate-density plot. 
In particular, the median rate of the compressing gas near the mean density becomes lower than in run \textsc{GT}. 
This suggests that magnetic fields serve to dampen the effects of turbulence on compressing gas. 
Furthermore, in regions where magnetic pressure dominates over self-gravity it will take longer for gas to move to higher density, which may explain the larger spread in the rates of low density gas relative to run \textsc{GT}.
At high densities, the expanding gas follows the free-fall rate closely, suggesting that magnetic pressure is sub-dominant to self-gravity above $s_{\mathrm{t}}$. 

In the final row of Fig.~\ref{fig:rate_dens_sum}, we see that the inclusion of protostellar outflows has an impact on the lowest-density gas.
This is in agreement with the findings of \cite{Appel+2022}, which show that protostellar outflows produce an excess of low-density gas.
The plots that include protostellar outflows show the same evidence of the mean flattening at densities near the mean density as in run \textsc{GTM}. 
However, run \textsc{GTMJR} also shows a large upturn in the rate of the very lowest-density gas -- for both compressing and expanding gas. 
This suggests that protostellar outflows are producing both rapidly expanding and rapidly compressing low-density gas. 
The compressing low-density material is likely associated to the bow shocks produced by the jets as they propagate through and entrain low-density gas.

\subsection{Net Compression and Expansion Rates} \label{sec:time}

\begin{figure*}[htb!]
\centering
\includegraphics[width = \linewidth]{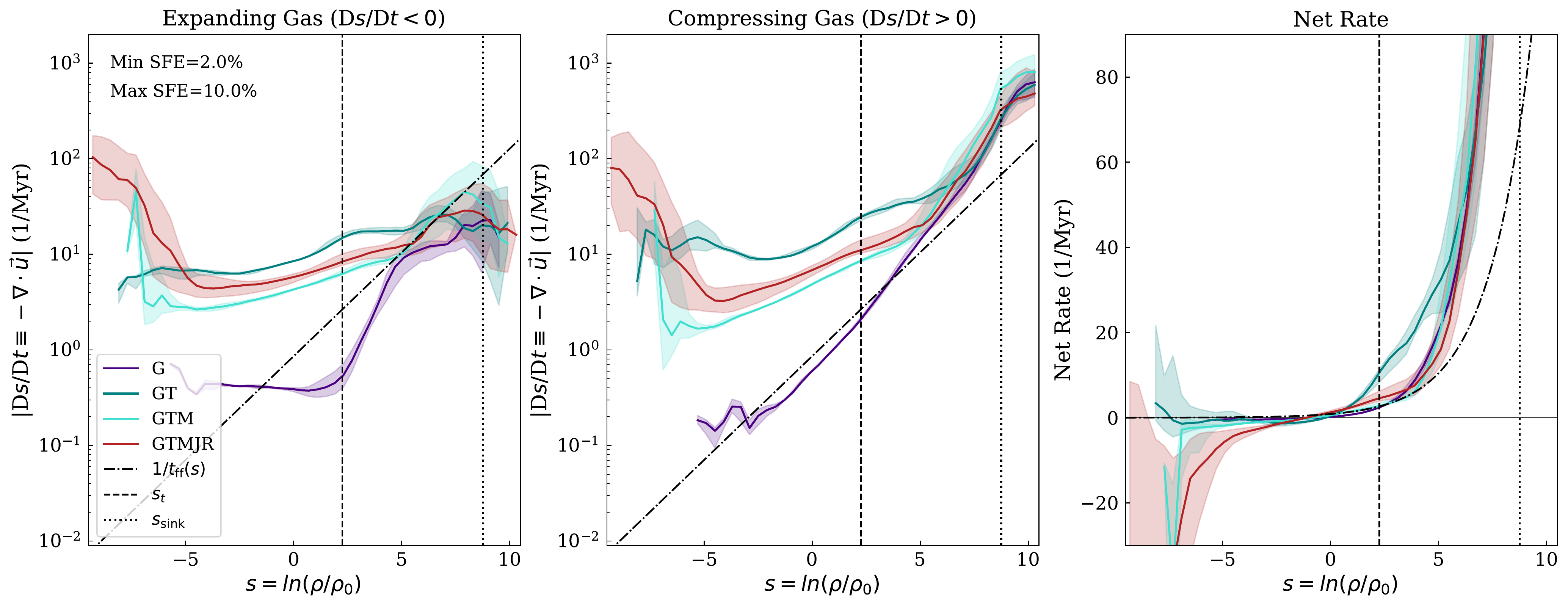}
\caption{The median in time ($\pm$ 1-sigma variation in time) of the median rates for the expanding gas ($Ds/Dt < 0$; left panel) and for the compressing gas ($Ds/Dt > 0$; center panel). 
The right-most panel shows the median in time ($\pm$ one sigma variation in time) of the net rate on a linear scale. The free-fall rate as a function of density is overplotted in each panel, as is the transition
density from \cite{Appel+2022} (vertical dashed line). The sink formation density ($s_{\mathrm{sink}}$) for each of the simulations is also shown as a dotted vertical line.
\label{fig:medians-net}}
\end{figure*}

In Fig.~\ref{fig:medians-net} we show the median compression and expansion rates as in Fig.~\ref{fig:rate_dens_sum}, but now averaged over time.
We take the median in time for snapshots between SFE$=2\%$ and SFE$=10\%$ in order to avoid fluctuations in the compression and expansion rates that we see at early times. 
These fluctuations at early times likely reflect the fact that it takes a while for the simulations to settle into an approximate steady state, and we exclude this transient period from our analysis.
All four simulations are shown with the 1-sigma variations shown as shaded regions around the median values.
The leftmost panel shows the median rate for the expanding gas and the central panel shows the median rate for the compressing gas. The rightmost panel of Fig.~\ref{fig:medians-net} shows the net rate, which we calculate by taking the volume-weighted average of all of the expanding gas rates and the compressing gas rates for a single density bin for a single snapshot. 
We then plot the median in time of this net rate in the rightmost panel of Fig.~\ref{fig:medians-net}.

Similar to the median lines for individual snapshots seen in Fig.~\ref{fig:rate_dens_sum}, we see that distributions of the median rates in Fig.~\ref{fig:medians-net} vary with the inclusion of additional physical processes.
The gravity-only simulation (run \textsc{G}) has a very low and flat distribution of expanding gas for densities around the mean density, in accordance with the lack of any turbulent velocities.
The net rate for run \textsc{G} increases with density, close to the free-fall rate for all densities, especially at low densities. 

The other simulations, which all include turbulent velocities and driving, have generally flatter (i.e., almost independent of density) distributions of rates around the mean density. 
Importantly, the turbulence-driven rates of gas expansion and compression at these densities ($s<5$; see Figs.~\ref{fig:rate_dens_sum}~and~\ref{fig:medians-net}) have comparable magnitudes and are much higher than the free-fall rate at the same densities (dashed-dotted line). 
This statistical equilibrium between compression and expansion is indicative of continuous gas cycling between the self-gravitating, high-density gas in the power-law tail, out of which stars form, and the low-density, non-star-forming gas that corresponds to the lognormal portion of the density PDF \cite[][see also \citealt{Semenov2017,Semenov2018} for analogous processes in the galactic context]{Appel+2022}.

At high densities, we see that the rate of the compressing gas for all simulations increases with density faster than the free-fall rate at densities well above $s_{\mathrm{t}}$ (e.g., $s\gtrsim 5$). 
The fact that all of the simulations converge on the same behavior as run \textsc{G} suggests that the compressing gas at high densities is strongly influenced by gravity in all cases.
The faster-than-free-fall collapse is likely a consequence of the details of the density distribution and dynamics of the gas that is accreting onto the sink particles. 
In particular, the faster-than-free-fall collapse may be a consequence of plotting the rate as a function of the $s$ value of individual cells; since $s$ is a cell-by-cell quantity, it may not reflect how the gas is actually distributed around the sinks and the resulting gravitational potential. 
Alternatively, this may indicate the influence of accretion processes.
In particular, previous work suggests that $s\sim5$ (which is where the net rate increases above free-fall) is where the accretion disk forms around the sinks \citep[i.e.,][]{Khullar2021}. 
This is also the density where the second power law forms in the density PDF \citep{FederrathKlessen2013,Burkhart2018,Khullar2021}.

In the net rate panel, we also see that for densities above the mean density, all of the physics cases are dominated by compressing gas. 
Only at densities below the mean does the net rate take on a negative value for some of the physics cases, corresponding to the gas being dominated by expansion. 
This effect is most evident in the net rate of the run with feedback (run \textsc{GTMJR}), which takes on large negative values for the lowest density gas (e.g., $s\lesssim -5$) as a result of the inclusion of protostellar jet feedback.
Similarly, at the lowest densities, we see that both the expanding and compressing gas rises significantly for run \textsc{GTMJR}.
This agrees with our understanding that jets drive gas out of dense regions and into low-density, rapidly expanding gas \citep[see also][]{Appel+2022}.

\subsection{The Gas Mass Flux} \label{sec:sfr_comparison}

\begin{figure*}[htb!]
\centering
\includegraphics[width = 0.75 \linewidth]{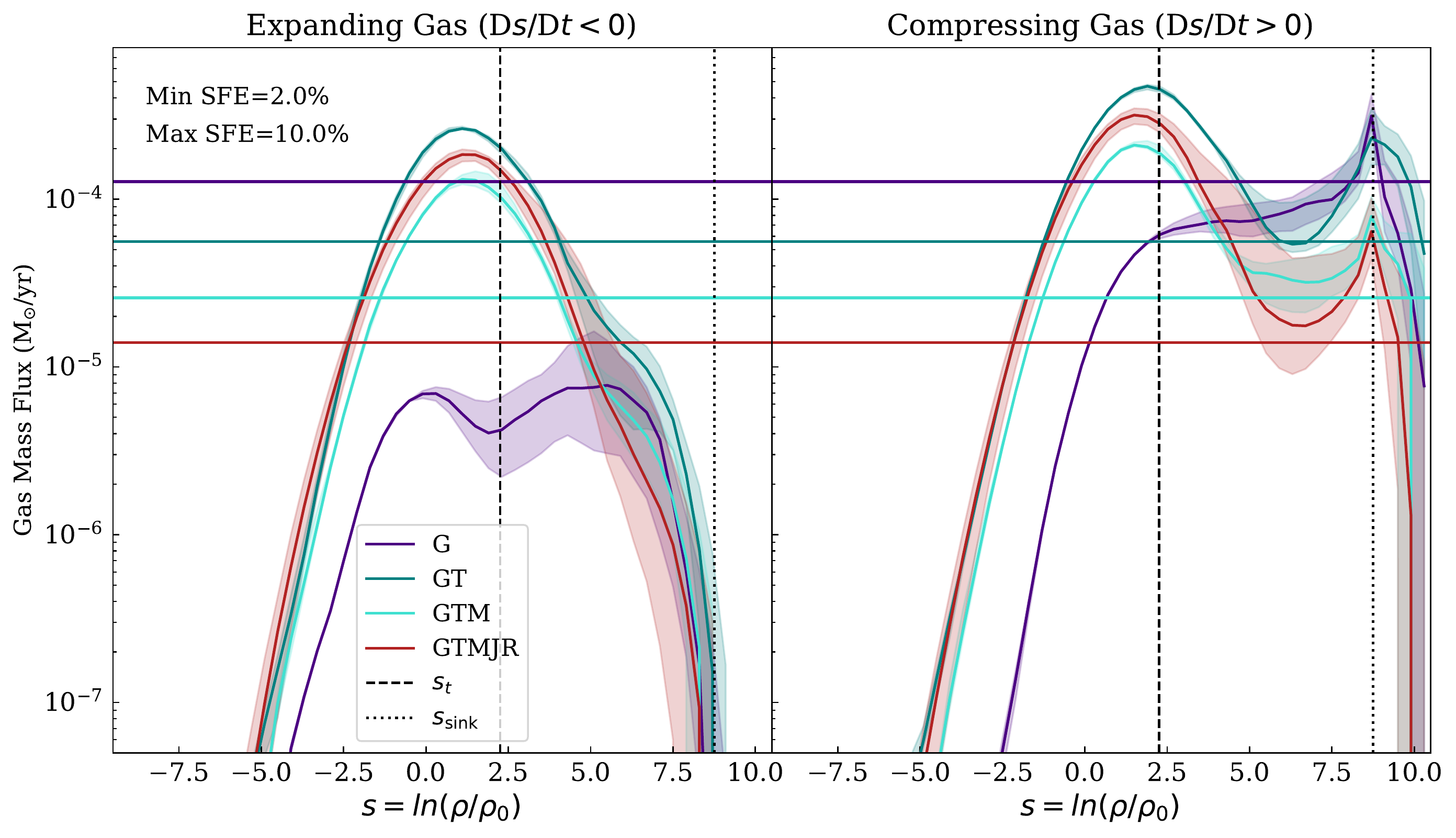}
\caption{The median in time ($\pm$ 1-sigma variation in time) of the gas mass flux in units of solar masses per year for the expanding gas ($Ds/Dt < 0$; left panel) and for the compressing gas ($Ds/Dt > 0$; right panel). The average
SFR of each run is overplotted as a horizontal
line. As in Fig.~\ref{fig:medians-net}, the transition
density from \cite{Appel+2022} and the sink formation density ($s_{\mathrm{sink}}$) for each of the simulations are also shown.
\label{fig:median-sfr}}
\end{figure*}

\begin{figure}[htb!]
\centering
\includegraphics[width = \linewidth]{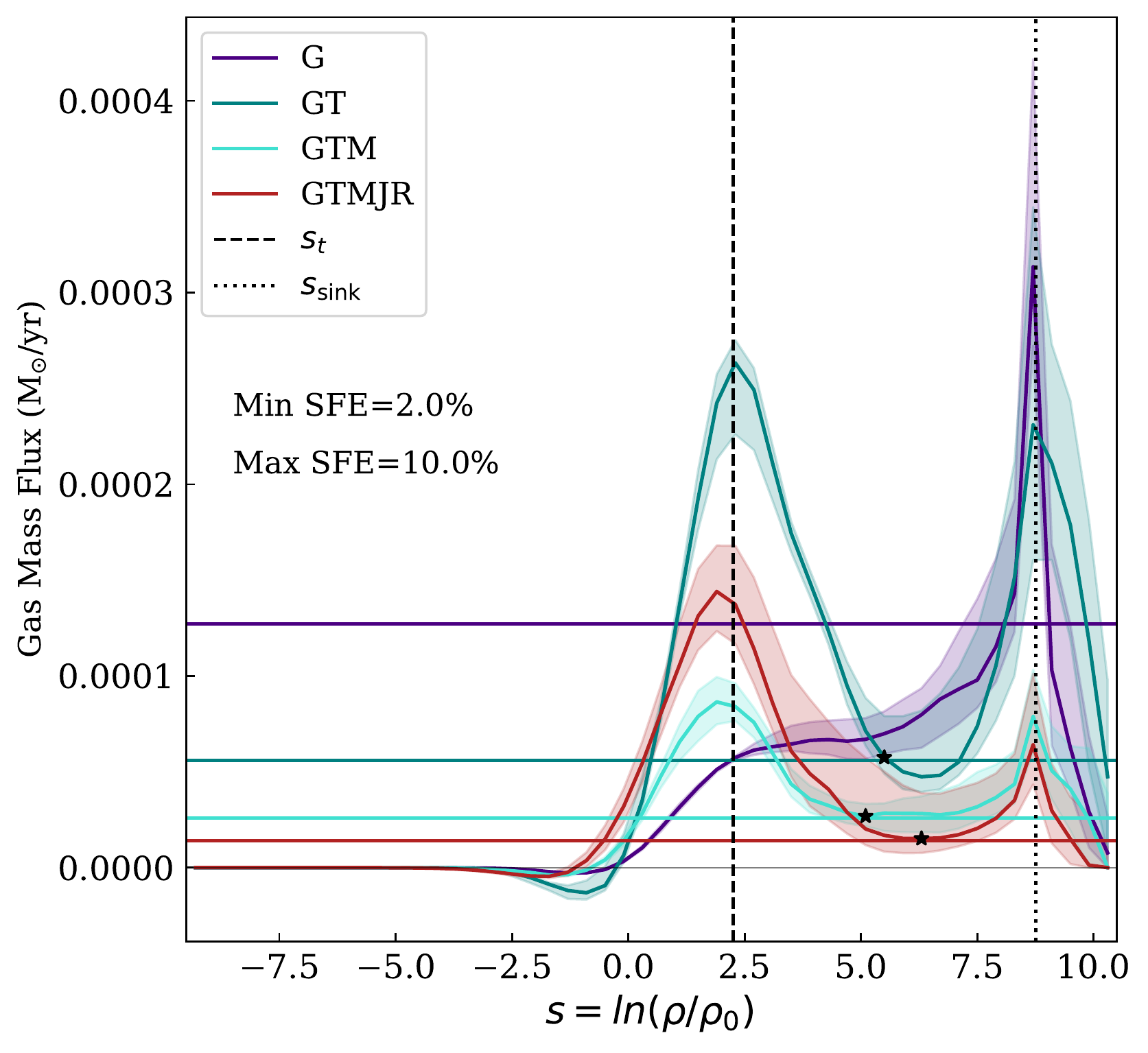}
\caption{The median in time ($\pm$ 1-sigma variation in time) of the net gas mass flux. 
The average SFR of each run is overplotted as a horizontal line. The values of $s_{*}$, the densities at which the net gas mass flux meets the SFR, for each of the runs except run \textsc{G} are shown as black stars. 
The horizontal, thin, grey line shows the SFR~$=0$ line. As in Fig.~\ref{fig:medians-net}, the transition density from \cite{Appel+2022} and the sink formation density ($s_{\mathrm{sink}}$) for each of the simulations are also shown.
\label{fig:net-flux}}
\end{figure}

In the previous sections, we considered the compression and expansion rates, which have dimensions of inverse time.  
By converting to a mass-weighted distribution of the rates versus density, we can calculate a gas mass flux in units of M$_{\odot}$/yr that we can compare directly to the SFR.
 
Let $H_m$ be the mass-weighted version of the heatmaps in Figs.~\ref{fig:rate_dens_all}~and~\ref{fig:rate_dens_sum}, or equivalently, the amount of mass at a given compression (or expansion) rate bin, $\Delta \mathcal{R}$, and density bin, $\Delta s$.
Then, the net gas mass flux (in  M$_{\odot}$/yr) for a given density bin is the product of $H_m$ and the compression rate, summed over every rate: 
\begin{equation}
    F_{\Delta s} = \sum_{\Delta \mathcal{R}} \, H_m \Delta \mathcal{R}\ \ .
\end{equation}
This gives us a net gas mass flux as a function of density, or a metric of how much gas is expanding or compressing in M$_{\odot}$/yr as a function of density. 
As with the net rate in Section~\ref{sec:time}, a negative net gas mass flux corresponds to net expanding gas and, conversely, a positive net gas mass flux corresponds to net compressing gas.

To get the gas mass flux for only the compressing gas, we sum over all of the gas with positive rates. 
Similarly, to get the gas mass flux for only the expanding gas, we sum over all of the gas with negative rates.

We show the median in time of the gas mass flux for the expanding gas (left panel) and the compressing gas (right panel) in Fig.~\ref{fig:median-sfr}, where the shaded regions show the 1-sigma variations in time.
The net gas mass flux is shown in Fig.~\ref{fig:net-flux}. 
For both Figs.~\ref{fig:median-sfr}~and~\ref{fig:net-flux}, we again have only considered snapshots between SFE=$2\%$ and SFE=$10\%$. 
We overplot the mean SFR (also for the snapshots between SFE=$2\%$ and SFE=$10\%$) for each simulation as a horizontal line in both figures.

\subsubsection{The Compressing and Expanding Gas Mass Flux}

The fluxes of both compressing and expanding gas show prominent peaks near the mean density, $s=0$. 
This is because there is much greater gas mass near $s=0$, where the mass-weighted density PDF peaks, and the fluxes quickly fall off at higher and lower densities where the density PDF also falls off. 
Compression clearly dominates near $s_{\mathrm{t}}$, which corresponds to the postshock density where mass piles up due to shocks \citep[][]{Federrath2016}.
At even higher densities, the expansion rate exponentially drops off. 
At $s\approx 6$ the compressing gas mass flux flattens out before rising again at yet higher densities. 
As Fig.~\ref{fig:net-flux} shows, this is also near the density at which the net gas mass flux is roughly equal to the SFR (see a more detailed discussion below). 

The compressing gas mass flux continues to increase above this density ($s\approx6$) and peaks at $s_{\mathrm{sink}}$.
We investigated the formation of this second peak and found that it develops as sink particles start forming. 
In particular, right before the first sink particle is formed, no peak is present at $s_{\mathrm{sink}}$, and the plateau in the compressing gas mass flux extends all the way to $s_{\mathrm{sink}}$ (except for run \textsc{G} which exhibits very different behavior, as discussed below). 
Therefore, this second peak is likely a numerical artifact resulting from the effects of limited resolution and of the sink particle model on the local gas dynamics.

Run \textsc{G} exhibits rather different behavior as a function of density and over time than the other runs. 
The median line shown in Figs.~\ref{fig:median-sfr}~and~\ref{fig:net-flux} for run \textsc{G} has roughly constant compressing gas mass flux with increasing density above $s_{\mathrm{t}}$ up until $s \approx 6$, after which the compressing (and net) gas mass flux increases and peaks at $S_{\mathrm{sink}}$. 
Unlike the other runs, however, before the peak at $s_{\mathrm{sink}}$ forms, there is not a plateau in the compressing gas mass flux at $s\approx6$ for run \textsc{G}; instead the compressing gas mass flux continues to decrease with increasing density above $s_{\mathrm{t}}$ until well after the first sink forms.
Indeed, the peak in the compressing gas mass flux at $s_{\mathrm{sink}}$ for run \textsc{G} develops slowly over many snapshots after the first sink is formed, suggesting this increase in compressing gas mass flux is due to the presence of sink particles. 
Furthermore, the lack of plateau before the formation of the first sink particle confirms that this plateau in the compressing gas mass flux is a consequence of physical processes beyond gravity.

We also note that the density range at which the compressing gas mass flux begins to increase again for all of the runs ($s\approx6$) corresponds to approximately the density at which the second power-law tail is expected to form in the density PDF, due to accretion disks beginning to form \citep[see e.g.,][]{Khullar2021}. 
Thus, the increase in flux just below the sink threshold density may also be influenced by the process of accretion onto sink particles.

The gas mass flux does fall off above $s_{\mathrm{sink}}$, but as there is only a very small amount of gas at these densities and this is, by definition, above the density at which sinks form, it is unclear how much we can trust any metrics of the behavior of the gas at these densities. 

\subsubsection{The Net Gas Mass Flux}

Looking at Fig.~\ref{fig:net-flux}, we see that the net gas mass flux is very close to zero at the lowest densities, corresponding to equal expanding and compressing gas mass flux, as expected for driven turbulence. 
At around the mean density, the net gas mass flux transitions rapidly to a positive value, indicating that gas at and above the mean density is net compressing. 
The transition to net compressing gas at approximately the mean density matches the behavior of the net rate in Fig.~\ref{fig:medians-net}, and is due to shock compression which piles up gas from the mean density to over-densities, up to the isothermal jump-condition of $M_s^2$ \citep[see e.g.,][and references therein]{Padoan11a,Federrath2016}.
The result of this shock compression is that the net gas mass flux peaks at the transition density, $s_{\mathrm{t}}$, indicating a gas pileup and the formation of filamentary structures in the cloud.

Similar to the behavior of the compressing gas mass flux in Fig.~\ref{fig:median-sfr}, the net gas mass flux decreases in all runs at densities greater than $s_{\mathrm{t}}$, except run \textsc{G} (the purple line in Fig.~\ref{fig:net-flux}). 
As discussed, for run \textsc{G}, gravity is the only dynamical process controlling the dynamics of the gas past $s_{\mathrm{t}}$ until sinks form and therefore the gas is moving with constant acceleration, hence the flat or increasing net gas mass flux between $s=s_{\mathrm{t}}$ and $s\approx5$). 
For the other simulations, the addition of turbulence dramatically changes the way gas moves around and past the post-shock density (i.e., the gas in the power-law portion of the density PDF). 
With driven supersonic turbulence, a strong peak at $s_{\mathrm{t}}$ in the gas mass flux confirms this density traces the post-shock density and the formation of filamentary features in the simulations \citep[see e.g.,][]{Padoan11a,Federrath2016}. 
At $s>s_{\mathrm{t}}$, gravity begins to dominate the dynamics, the density PDF forms the first power-law tail, and the gas mass flux falls off until it reaches a plateau at around $s=5-6$ (marked with a star symbol in Figure ~\ref{fig:net-flux}). 

For ease of reference, we refer to the density at which the net gas mass flux first matches the mean SFR (after rising above the SFR at $s_{\mathrm{t}}$) as $s_{*}$. 
The values of $s_{*}$ for each of the runs except run \textsc{G} (which exhibits a very different behavior between $s_{\mathrm{t}}$ and $s_{\mathrm{sink}}$, as discussed) are reported in Table~\ref{tab:sims} and are shown as black stars in Fig.~\ref{fig:net-flux}. 
The values in Table~\ref{tab:sims} and Fig.~\ref{fig:net-flux} are calculated for the mean SFR and median gas mass flux between SFE=$2\%$ and SFE=$10\%$.

The fall off in the gas mass flux between $s_{\mathrm{t}}$ and $s_{*}$ is due to the fact that the acceleration (i.e., derivative of the rate) is smaller than free-fall collapse would suggest due to magnetic pressure and turbulent support, thereby slowing down collapse. 
The acceleration picks up at around $s=5-7$, where an increase in the slope of the compression rate, shown in the middle panel of Fig.~\ref{fig:medians-net}, can be observed. 
At this point, the plateau develops, signaling a constant gas mass flux. 
Interestingly, the value of the gas mass flux at this plateau matches the mean SFR.
At the highest densities, a strong peak develops around the sink threshold density, as mass is rapidly funneled into the sink particles. 

\begin{figure}[tb]
\centering
\includegraphics[width = 0.9 \linewidth]{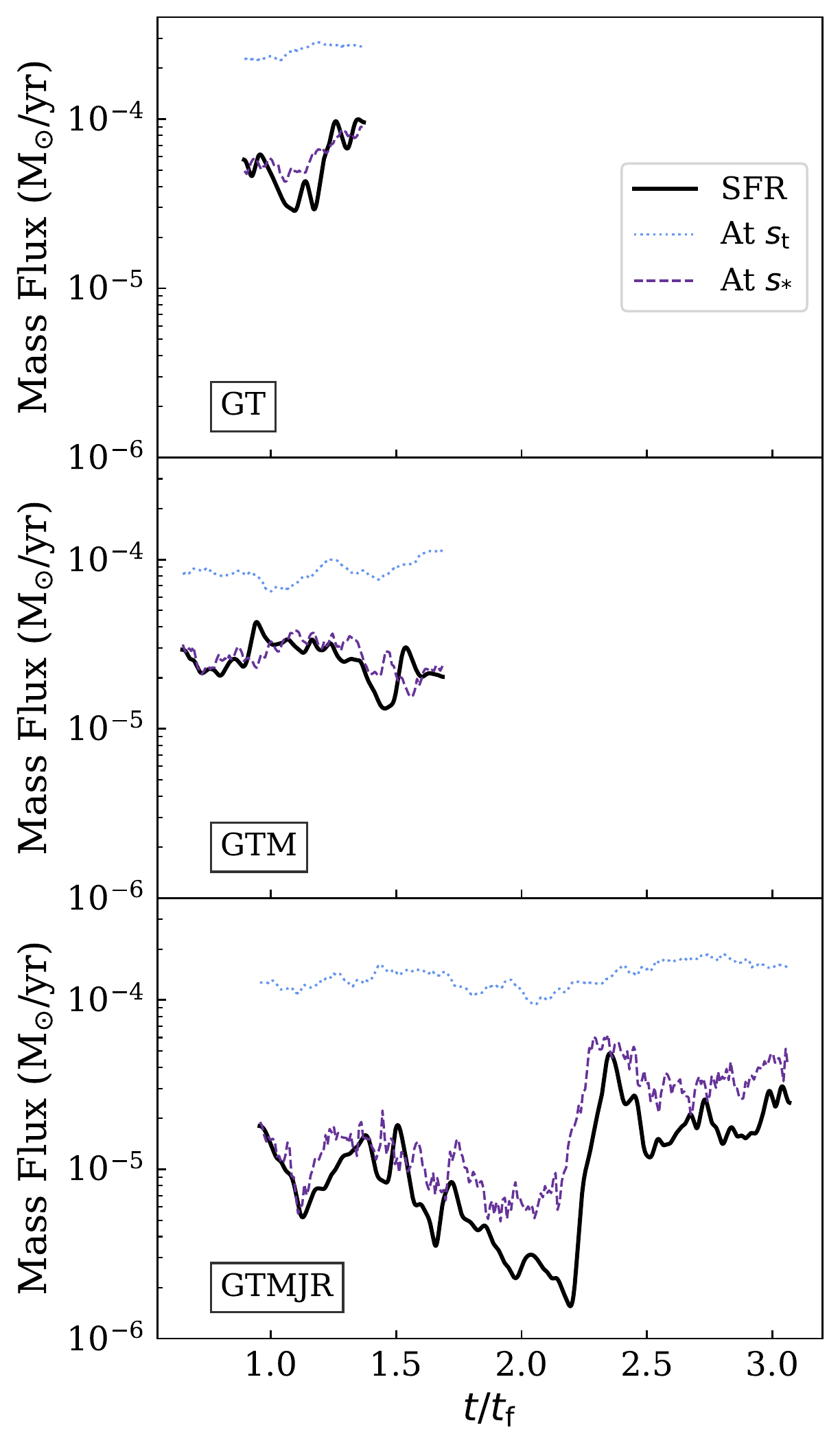}
\caption{The net gas mass flux at two different fixed densities ($s = s_{\mathrm{t}}$ and $s = s_{*}$) are plotted in comparison to the smoothed SFR as a function of time for each simulation. The net gas mass flux values are the time dependent counterparts to the values in Fig.~\ref{fig:net-flux} and are measured at a single density bin with a center equal to or just above the corresponding density.
\label{fig:sfr_v_t}}
\end{figure}

\subsubsection{Connecting the Gas Mass Flux and the SFR}

We further investigate the relationship between $s_{*}$, the gas mass flux, and the SFR in Fig.~\ref{fig:sfr_v_t}, where we plot the SFR as a function of time for runs \textsc{GT}, \textsc{GTM}, and \textsc{GTMJR}. 
The SFR is smoothed using \verb|scipy|'s \verb|gaussian_filter1d| function with a sigma of $\sim 0.02$~Myr. 
We plot the net gas mass flux value (sampled at every fifth snapshot to reduce noise) at two key density bins: at $s_{\mathrm{t}}$ and at $s_{*}$. 
The values plotted in Fig.~\ref{fig:sfr_v_t} are the value of the net gas mass flux for a single snapshot at the density bin at or just above the density of $s_{\mathrm{t}}$ or $s_{*}$. 
We note that we use the same value of $s_{*}$ (the value in Table~\ref{tab:sims}) for every snapshot, and, since this value is calculated based on where the median net gas mass flux meets the mean SFR, may not reflect exactly where the net gas mass flux of an individual snapshot meets the SFR of that snapshot. 

We find that the net gas mass flux at the transition density, $s_{\mathrm{t}}$, is higher than the SFR by about $40\%$. 
This is consistent with Fig.~\ref{fig:net-flux} which shows that the median gas mass flux near $s_{\mathrm{t}}$ is higher than the mean SFR for all three cases.
This discrepancy likely corresponds to the core mass efficiency factor for star formation models which take all the gas in the power-law tail portion of the density PDF as star-forming \citep{Burkhart2018}.
The core mass efficiency factor accounts for the fact that the inclusion of turbulence and magnetic fields makes the process of forming stars take much longer than free-fall and that the inclusion of outflows cycles dense gas back to densities below $s_{\mathrm{t}}$ before it forms stars. 
This slow down of the formation of stars from dense, star-forming gas is clearly evident in the discrepancy between the net gas mass flux at $s_{\mathrm{t}}$ and the smoothed SFR.
Furthermore, although the difference between the net gas mass flux at $s_{\mathrm{t}}$ and the SFR is largest (at most times) for run \textsc{GTMJR}, there is a significant discrepancy for all three runs shown in Fig.~\ref{fig:sfr_v_t}, confirming that both turbulence and magnetic fields, and not only outflows, are contributing to preventing gas above the transition density from forming sink particles.

However, for all three runs, the net gas mass flux at $s_{*}$ as a function of time closely matches the value of the SFR. 
Time variable features of the SFR are also found in the net gas mass flux at $s_{*}$. 
For example, there is a large jump in the SFR for run \textsc{GTMJR} around $t/t_{\mathrm{ff}} \sim 2.2$ that is echoed by a similar jump in the net gas mass flux at $s_{*}$, but not in the gas mass flux at $s_{\mathrm{t}}$. 
The similarity between the value of the net gas mass flux at $s_{*}$ and the SFR indicates that the gas dynamics around $s_{*}$ play a key role in setting the SFR. 
We explore some of the implications of this connection in the Discussion section below.

\section{Discussion} \label{sec:discussion}

\subsection{The Compression and Expansion Rates}

Our work investigates the ways in which different physical processes affect the gas dynamics of star-forming regions as a function of density and how they are reflected in the density PDF shape. 
In density regimes where the influence of gravity is dynamically dominant (e.g., above the transition density $s_{\mathrm{t}}$), the gas dynamics of the simulations that include turbulence do not completely match the behavior of the gravity-only run until densities above $s \gtrsim 5$ (i.e., at densities where self-gravity dominates).
Only at the highest densities of compressing gas does the behavior of the compression rate become very similar for all of the simulations presented in this work. 
This suggests that turbulence, magnetic fields, and feedback act to significantly alter how much gas reaches the highest densities and influence the structure of collapsing regions.

Similarly, we see that turbulence acts to increase the rate of both the compressing and expanding gas at densities below the transition density. 
Below the transition density turbulence dominates, and the compressing and expanding motions balance out, resulting in a net rate that is near zero. 
Above the transition density, gravity dominates and the net rate rises rapidly. 
We also find that the net gas mass flux peaks at $s_{\mathrm{t}}$, suggesting that the density where the first power-law tail forms is an excellent tracer of the post-shock density \citep[see e.g.,][]{Padoan11a,Federrath2016}. 

In contrast to the effect of turbulence, magnetic fields act to decrease the rate of both the compressing gas and the expanding gas at most densities below the transition density, relative to the run with only turbulence and gravity. 
This suggests that magnetic pressure acts to dampen the increased motion from turbulence. 

Finally, we see that the inclusion of protostellar outflows slightly increases the median rates of both the compressing gas and the expanding gas, relative to the run with magnetic fields (as seen in Fig.~\ref{fig:medians-net}), in agreement with the expectation that the inclusion of protostellar outflows will increase the kinetic energy of the gas \citep{Appel+2022}. 
The most dramatic effect of protostellar outflows is on the lowest density gas, where we see that protostellar outflows produce rapidly expanding and compressing low-density gas. 
In addition, the net rate (rightmost panel in Fig.~\ref{fig:medians-net}) has a lot more variation in time for the cases with protostellar outflows at low densities, suggesting that the inclusion of protostellar outflows introduces significant time variation in the compression and expansion rates of the low-density gas carved out by outflows.

In Fig.~\ref{fig:medians-net}, we find that the compression rate increases with density faster than the free-fall rate for all physics cases, once the density exceeds approximately $s \sim 5$. 
As discussed above, this may be a consequence of plotting the rate as a function of $s$, when $s$ is a cell-by-cell quantity and may not reflect how the gas is actually distributed or the resulting gravitational potential.

This density also heralds the formation of the second power-law tail in these simulations, as studied by \citet{Khullar2021} and which roughly corresponds to the formation of accretion disks. 
This process of accretion onto the sinks may also contribute to the faster than free-fall collapse seen in Fig.~\ref{fig:medians-net}, although our runs likely do not fully resolve accretion disks, making this connection uncertain.
Regardless, the rate at which the gas passes through this density range appears to play a role in setting the star formation rate in our simulations (see Figs.~\ref{fig:net-flux}~and~\ref{fig:sfr_v_t}, and discussed below). 
Future work will determine how the accretion disk forms and how the gas mass flux and compression and expansion rates depend on the sonic Mach number, Alfv\'enic Mach number, and virial parameters.

In Section~\ref{sec:rate-dens-plot} we also explored the relationship between the compression and expansion rates and the density PDF. 
We see that the high-density end of the density PDF is most closely matched by the PDF of the compressing gas and the low-density end of the density PDF is most closely matched by the PDF of the expanding gas.
This agrees with our understanding that the compressing gas is mostly at higher densities (that are dominated by gravity) and the expanding gas is mostly at lower densities where gravity is subdominant.

\subsection{The Gas Mass Flux}

The gas mass flux (Fig.~\ref{fig:median-sfr}) combines information from the compression and expansion rates (Fig.~\ref{fig:medians-net}) and the overall density PDF. 
At low densities, the gas mass flux is low due to both low rates and small quantities of gas. 
Near the transition density from lognormal to power-law distributions, the gas mass flux peaks due to the formation of shocks \citep{Federrath2016}.
The drop-off in the net gas mass flux above the transition density is evidence of various processes actively preventing collapse of the gas since the acceleration of the gas is stalled relative to free-fall acceleration.
The net gas mass flux declines until it reaches a constant value (i.e., it plateaus), analogous to a terminal velocity where the resistive forces are magnetic fields and turbulent motions.
The net gas mass flux at this plateau matches the SFR at a density ($s_*$) that is well above the transition density. 
This behavior of the gas mass flux matches the fact that analytical models of the SFR that integrate over all densities above the critical density require an additional efficiency factor, implying that not all of the gas above the critical density ends up in a star. 
This agrees with the fact that the gas mass flux at $s_{\mathrm{t}}$ is higher than the SFR before dropping off -- there are processes preventing and delaying much of this gas from actually forming stars.

\cite{Khullar2021} demonstrate the existence of a second power-law in the density PDF that begins at densities greater than $s\sim 5$. 
Their model suggests that the lognormal portion of the density PDF is turbulence dominated, the first power-law is gravity dominated, and the second power-law (corresponding to the highest density gas) is disk or rotation dominated. 
This suggests a possible interpretation for the behavior of the net gas mass flux at high densities. 
In particular, the point where the net gas mass flux matches the SFR ($s_*$) may correspond to the beginning of this second power-law and the increase in the net gas mass flux above $s_*$ may be due to the influence of disk rotation. 
Although, again, our simulations do not fully resolve the disk accretion, meaning that further work is needed to verify this connection. 
As discussed above, however, the flux at which the gas passes through this density range appears to set the star formation rate in our simulations.
This rate is highest in run \textsc{GT} and lower in the runs with feedback from protostellar outflows; hence the SFR is lower when outflow feedback is included.  
Further work is needed to confirm this connection and to compare the $s_{\mathrm{d}}$ value from \cite{Khullar2021} to the value of $s_*$ found here.

\subsection{Other Implications and Future Work}

Our work may have important implications for sub-grid models for isolated GMC simulations, or even galaxy formation simulations.
We demonstrated that the SFR is set by the gas mass flux at $s_*$. 
This can act as a minimum resolvable density required to set the SFR in simulations. 
However, measuring the gas mass flux peak and fitting a curve to higher densities could result in an empirical sub-grid SFR model that could be used by simulations to ``resolve" protostellar core physics. 
Doing so would require measuring $\nabla \cdot \vec{v}$ and determining where this is negative (i.e., where gas is compressing). 
Turning this into a gas mass flux (Fig.~\ref{fig:median-sfr}) could then yield similar curves which could be extrapolated to higher-than-resolved densities (i.e., protoplanetary disk densities) where the star formation rate is then set. 

Future work will explore how this $s_{*}$ density depends on the cloud mass, the virial parameter, the sonic Mach number, the Alfv\'enic Mach number, and the magnetic field properties. 
We would also like to study cases without driven turbulence and with and without self-consistent feedback driven turbulence. 
Future studies could also explore how the gas dynamics, and $s_{*}$ in particular, changes with the inclusion of more realistic thermal physics, such as that associated with an ambient FUV field and comsic rays \citep[e.g., similar to the thermal physics set up in][]{Wu+2017}.
As discussed, further work is also needed to understand the potential connection between the values of $s_*$ and the $s_{\mathrm{d}}$ value from \cite{Khullar2021}, as well as the role of disk rotation and accretion.

\section{Conclusions} \label{sec:conclusion}

Previous work has presented evidence for the cycling of gas between different parts of gas density PDF within molecular regions: the high-density power-law tail, out of which stars form, and the non-star-forming log-normal portion at average and low densities \citep{Appel+2022}.
In this paper, we build on this analysis and further investigate the gas dynamics within star-forming regions using metrics such as the compression and expansion rates of the gas as a function of the gas density, and the gas mass flux through different portions of the density PDF. 

We find that:
\begin{itemize}
     \item The overall gas dynamics are dominated by compressing gas at densities above the mean density (corresponding to the power-law part of the density PDF), in agreement with the fact that the simulations are undergoing net gravitational collapse at high densities. In particular, at the highest densities, the net rate of all of our runs matches the net rate of the run with only gravity, suggesting that processes other than gravity have little effect at these densities.
    \item At average to low densities (corresponding to the lognormal part of the density PDF), turbulence produces both compression and expansion, and results in a relatively constant rate, independent of gas density. This rate is significantly higher than the free-fall rate at these low densities.
    \item We find that the net gas mass flux peaks at the transition between the lognormal and power-law forms of the density probability distribution function. This is consistent with the transition density tracking the post-shock density, which promotes an enhancement of mass at this density (i.e., shock compression and filament formation).
    \item The inclusion of stellar feedback in the form of protostellar outflows has a significant effect on the gas dynamics at low densities where protostellar outflows result in very rapidly expanding and compressing gas.
    \item For simulations that include turbulent velocities, the net gas mass flux above the transition density declines until it reaches a constant value (i.e., it plateaus).
    The net gas mass flux becomes constant at a density within the power-law tail, which we denote as $s_*$. The gas mass flux at $s_*$ closely traces the SFR, despite it being a far lower density than the sink threshold.
    This suggests that the gas dynamics at this density, $s_*$, play an important role in setting the SFR. We find that $s_*$ varies slightly with the inclusion of different physics.
\end{itemize}

\acknowledgments  

S.M.A. \& B.B. acknowledge support from NSF grant AST-2009679.
B.B. is grateful for generous support by the David and Lucile Packard Foundation and Alfred P. Sloan Foundation. 
Support for V.S. was provided by NASA through the NASA Hubble Fellowship grant HST-HF2-51445.001-A awarded by the Space Telescope Science Institute, which is operated by the Association of Universities for Research in Astronomy, Inc., for NASA, under contract NAS5-26555, and by Harvard University through the Institute for Theory and Computation Fellowship. 
C.F.~acknowledges funding by the Australian Research Council (Future Fellowship FT180100495 and Discovery Projects DP230102280), and the Australia-Germany Joint Research Cooperation Scheme (UA-DAAD). A.L.R.~acknowledges support from the National Science Foundation (NSF) Astronomy and Astrophysics Postdoctoral Fellowship under award AST-2202249.
J.C.T. acknowledges support from NSF grant AST-2009674.
The analysis and simulations were performed using computing resources provided by the Flatiron Institute and the NCI Gadi cluster. 
We further acknowledge high-performance computing resources provided by the Leibniz Rechenzentrum and the Gauss Centre for Supercomputing (grants~pr32lo, pn73fi, and GCS Large-scale project~22542), and the Australian National Computational Infrastructure (grant~ek9) in the framework of the National Computational Merit Allocation Scheme and the ANU Merit Allocation Scheme.
The software used in this work was developed in part by the DOE NNSA - and DOE Office of Science - supported Flash Center for Computational Science at the University of Chicago and the University of Rochester.
The authors further acknowledge the use of the following software: yt \citep{Turk11a}, flash \citep{Fryxell2000}, SciPy \citep{SciPy}, scikit-learn \citep{scikit-learn}, Matplotlib \citep{Hunter2007}, astropy \citep{Astropy-Collaboration13a}. 

\appendix
\section{Eulerian and Lagrangian Continuity Equation} \label{sec:app_lagrangian}

In Section~\ref{sec:rate-method}, we use the Lagrangian formulation of the continuity equation. 
Here, we briefly show how to derive the Lagrangian formation of the continuity equation. 

First, let us consider a fluid element of volume $V$ and density $\rho$. 
The mass of this element remains constant in time, even as the density and volume may change. 
Thus,
\begin{align}
    \frac{D}{Dt} \left( \rho V \right) & = 0 
    \quad\iff\quad V \frac{D \rho}{D t} + \rho \frac{D V}{D t}  = 0
    \label{eq:exp_mass_cons_app}
\end{align}
where we use $D/Dt$ as a reminder that we are using the Lagrangian derivative.
Considering only the second term of the latter expression:
\begin{align}
    \frac{D V}{D t} & = \oint_{A} \left( \vec{v} \cdot \vec{A} \right) d s \nonumber  \\
    & = \int_{V} \left( \nabla \cdot \vec{v} \right) d V \nonumber \\
    & = \left( \nabla \cdot \vec{v} \right) V
\end{align}
where $\vec{v}$ is the velocity field, $\vec{A}$ is the normal vector of the surface, and we have used the Divergence theorem to go from the first to the second line. 
We can then rewrite Eq.~\ref{eq:exp_mass_cons_app} as:
\begin{align}
    V\frac{D \rho}{D t} + \rho \left[ \left( \nabla \cdot \vec{v} \right) V\right] & = 0 \nonumber \\
    \frac{D \rho}{D t} + \rho  \left( \nabla \cdot \vec{v} \right) &=0 
    \label{eq:conteq_rho_app}
\end{align}
which gives us the familiar Lagrangian formulation of the continuity equation in terms of $\rho$, shown in Eq.~\ref{eq:conteq_rho}.  

Since we wish to compare this expression to the density PDF, which we have calculated in terms of $s=\ln(\rho/\rho_0)$, we rearrange Eq.~\ref{eq:conteq_rho_app} in terms of $s$.
First, we rewrite $\rho = \rho_0 \, e^{s}$.
Then, 
\begin{align}
    \frac{D }{D t} \left( \rho_0 \, e^{s} \right)+ \left(\rho_0 \, e^{s} \right)  \left( \nabla \cdot \vec{v} \right) &=0 \nonumber \\
    \left( \rho_0 \, e^{s} \right)\frac{D s}{D t} + \left(\rho_0 \, e^{s} \right)  \left( \nabla \cdot \vec{v} \right) &=0 \nonumber \\
    \frac{D s}{D t} + \left( \nabla \cdot \vec{v} \right) &=0
\end{align}
Thus,
\begin{equation}
    \frac{D (s)}{Dt} \equiv - ( \nabla \cdot \Vec{v} \,)
\end{equation}
where $s = \ln (\rho/\rho_{0})$. 
Thus, we have a connection between the time evolution of the density (the $Ds/Dt$ term) and the gas dynamics (as represented by the velocity vector, $\Vec{v}$).

However, we can also derive the Lagrangian continuity equation from the Eulerian formulation. 
From the standard Eulerian continuity equation, we find
\begin{align}
    & \frac{\partial\rho}{\partial t} + \nabla\cdot\left(\rho\Vec{v}\right) = 0 \nonumber \\
    \iff\quad & \frac{\partial\rho}{\partial t} + \left(\Vec{v}\cdot\nabla\right)\rho = - \rho\nabla\cdot\Vec{v} \nonumber \\
    \iff\quad & \frac{1}{\rho}\frac{D\rho}{Dt} = - \nabla\cdot\Vec{v} \nonumber \\
    \iff\quad & \frac{Ds}{Dt} = - \nabla\cdot\Vec{v},
\end{align}
where the Lagrangian (co-moving) derivative $D/Dt = \partial/\partial t + (\Vec{v}\cdot\nabla)$ and $s = \ln (\rho/\rho_{0})$ were used in the last two steps.

\section{Analysis of Run  \textsc{GTMJ}} \label{sec:app_J}

\begin{figure*}[htb!]
\centering
\includegraphics[width = 0.9 \linewidth]{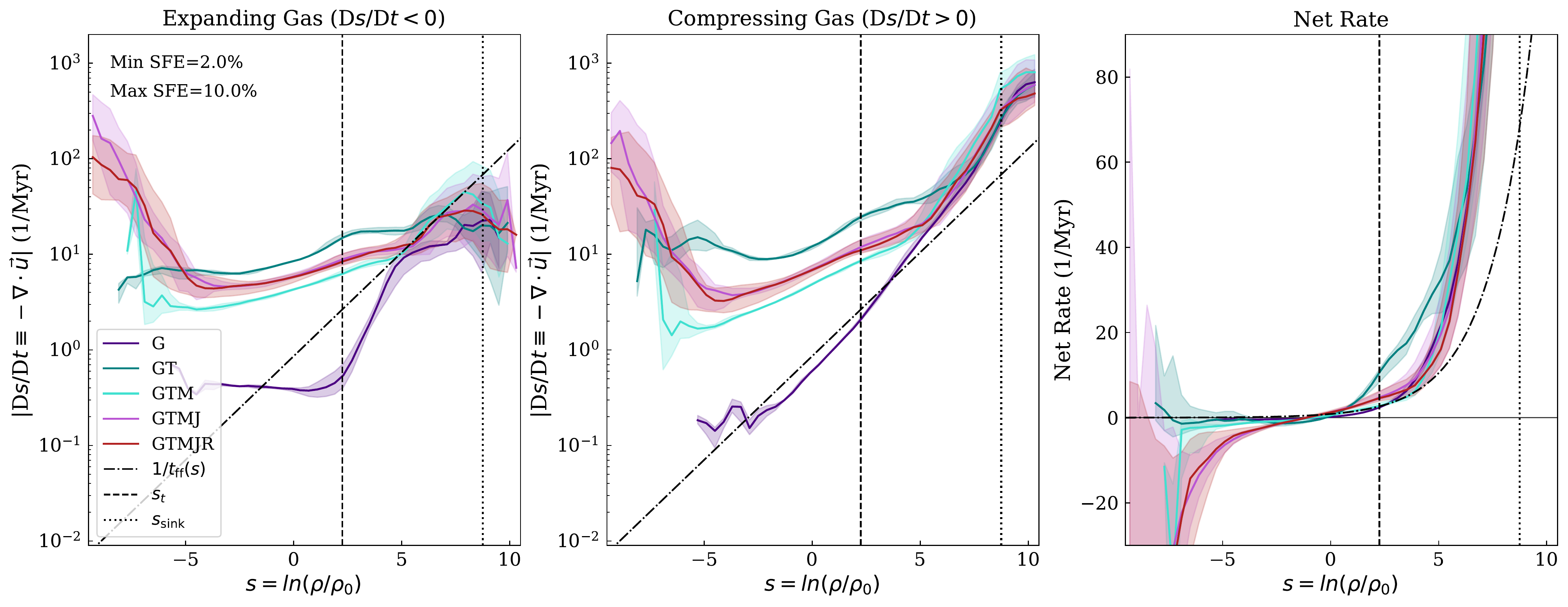}
\caption{Figure~\ref{fig:medians-net} with run \textsc{GTMJ}.
\label{fig:medians-net-withJ}}
\end{figure*}

\begin{figure*}[htb!]
\centering
\includegraphics[width = 0.7 \linewidth]{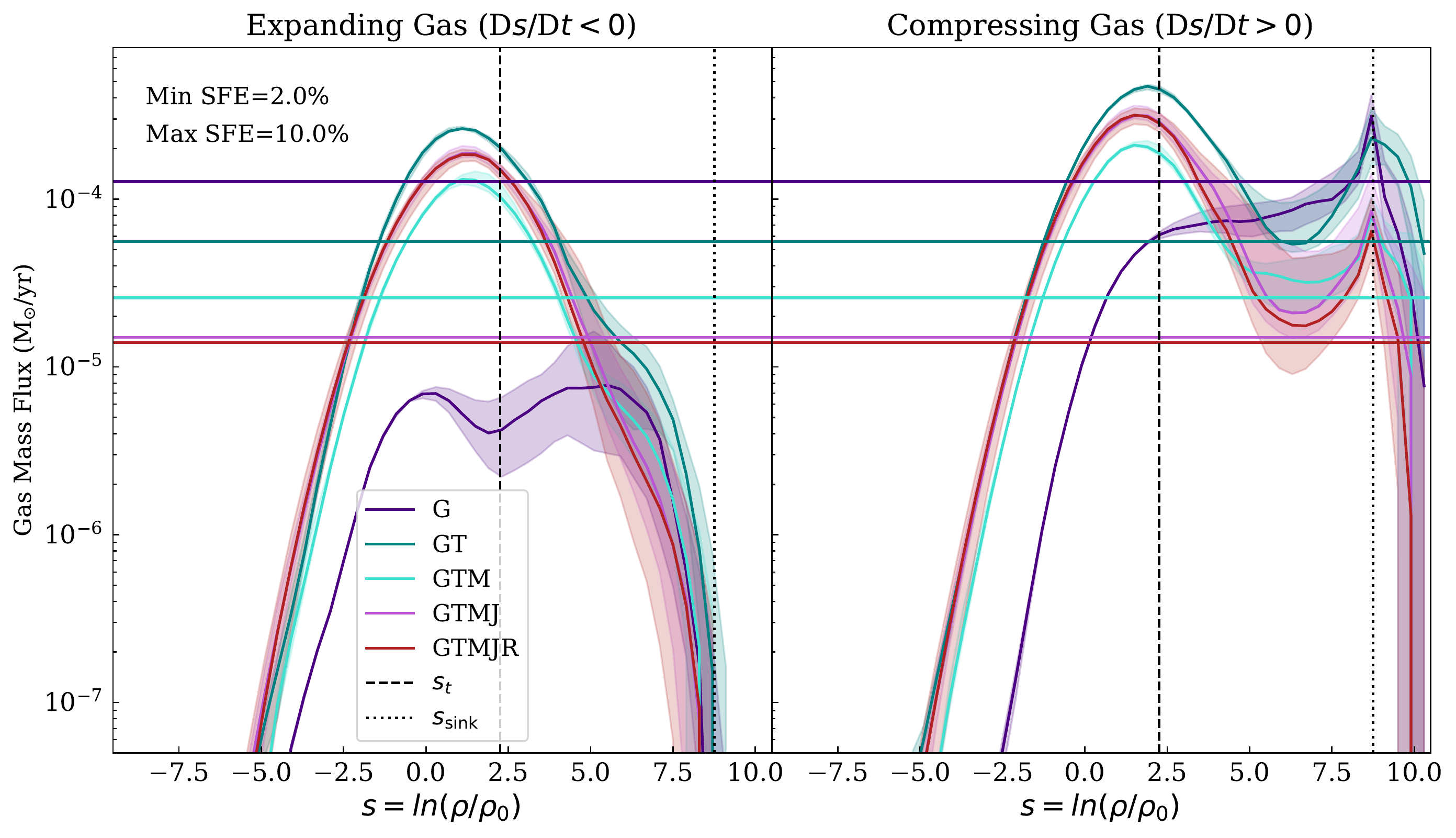}
\caption{Figure~\ref{fig:median-sfr} with run \textsc{GTMJ}.
\label{fig:median-sfr-withJ}}
\end{figure*}

\begin{figure}[htb!]
\centering
\includegraphics[width = \linewidth]{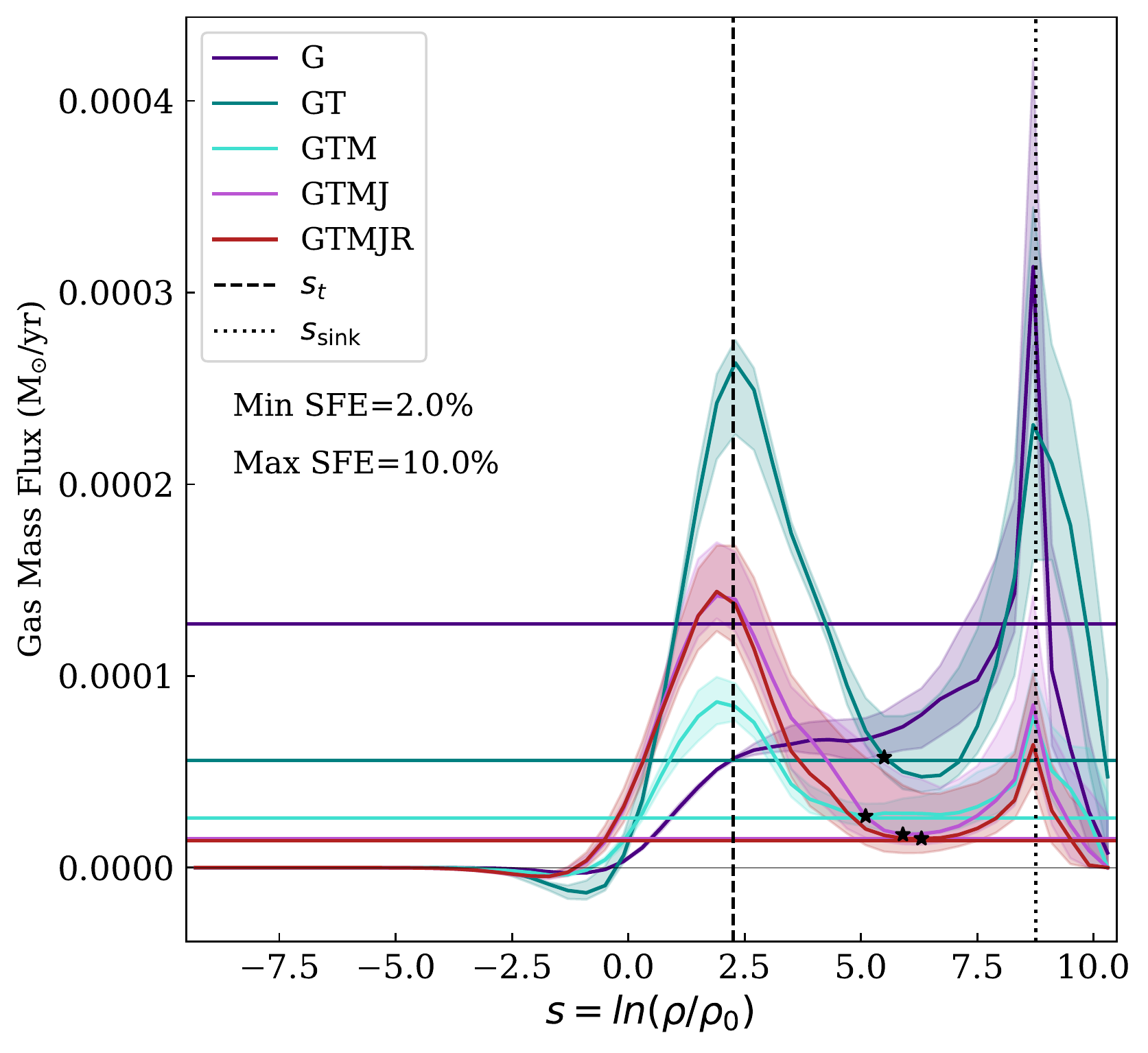}
\caption{Figure~\ref{fig:net-flux} with run \textsc{GTMJ}.
\label{fig:net-flux-withJ}}
\end{figure}

In Sections~\ref{sec:analysis} through~\ref{sec:discussion}, we focused on run \textsc{GTMJR} and did not show the results for run \textsc{GTMJ} since the differences between these runs for the purposes of our analysis are minimal. 
We found that including both lines in our figures significantly cluttered our plots without substantially enhancing the understanding of our results.
However, for completeness, we use this appendix to present Figs.~\ref{fig:medians-net},~\ref{fig:median-sfr},~and~\ref{fig:net-flux} with run \textsc{GTMJ} also shown. 

Figure~\ref{fig:medians-net-withJ} reproduces Fig.~\ref{fig:medians-net} with the addition of run \textsc{GTMJ}.  
Very little difference can be found between runs  \textsc{GTMJ} and \textsc{GTMJR}, although there seems to be slightly more time variation in run \textsc{GTMJR} for the net rate at the lowest densities.

Figure~\ref{fig:median-sfr-withJ} reproduces Fig.~\ref{fig:median-sfr} with the addition of run \textsc{GTMJ}. 
Again, there is very little difference between runs \textsc{GTMJ} and \textsc{GTMJR}. 
In fact, the mean SFRs are almost identical. 
The inclusion of radiative heating in run \textsc{GTMJR} appears to slightly lower the median value of the compressing gas mass flux relative to run \textsc{GTMJ} near $s\sim6.5$, however the difference is small and well within the 1-sigma time variation of both runs.

Figure~\ref{fig:net-flux-withJ} reproduces Fig.~\ref{fig:net-flux} with the addition of run \textsc{GTMJ}. 
Again, there is very little difference between runs \textsc{GTMJ} and \textsc{GTMJR}. 
Indeed, the value of $s_*$ is very similar for the two runs, as can be seen in Table~\ref{tab:sims}.

\section{Check time variation of the density PDF} \label{sec:app_PDFs}

\begin{figure}[htb!]
\centering
\includegraphics[width = \linewidth]{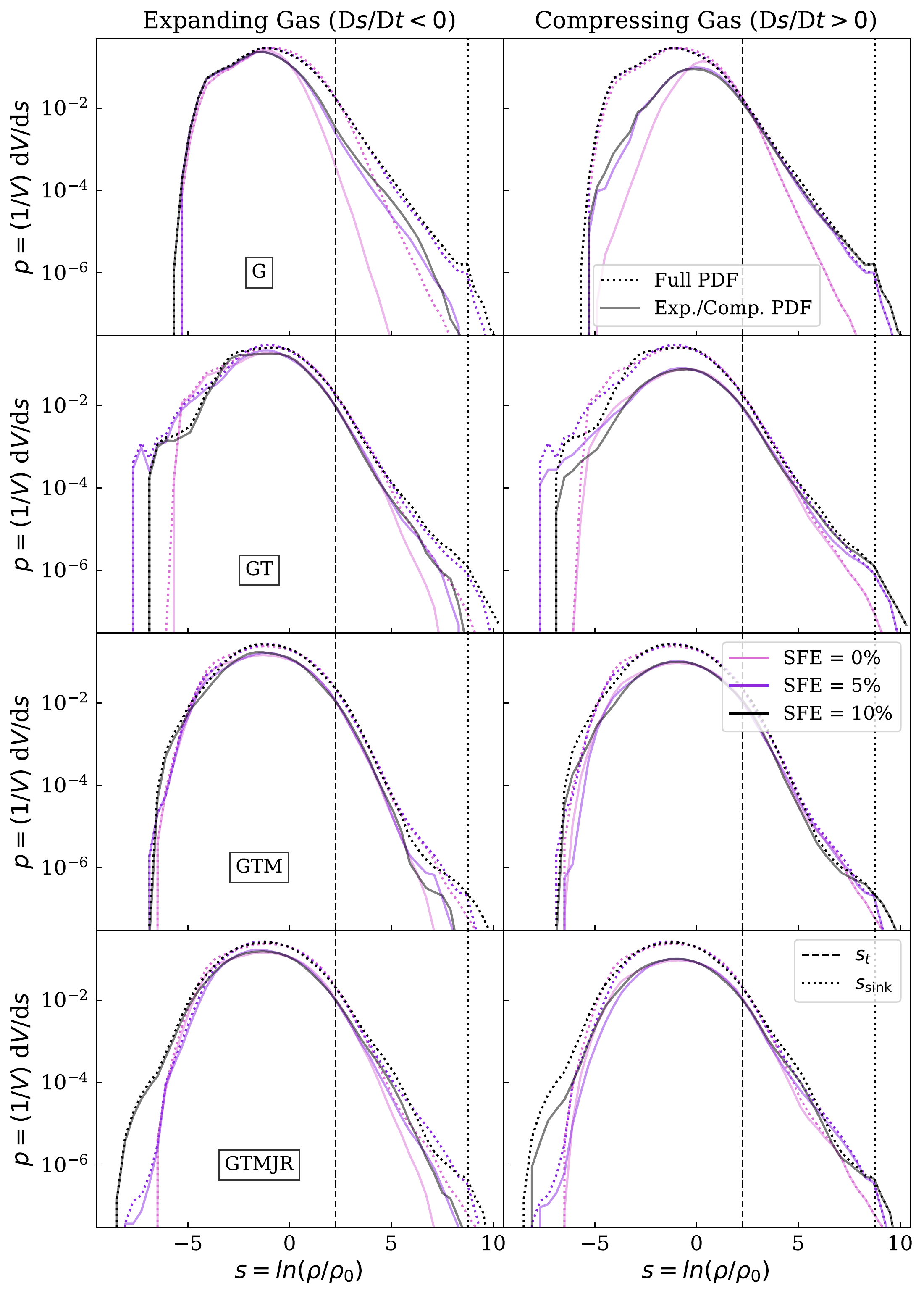}
\caption{Each panel shows the volume-weighted density PDF for all of the gas in the simulation region (dotted line) and the volume-weighted density PDF for only the expanding gas ($Ds/Dt < 0$; left column) or only the compressing gas ($Ds/Dt > 0$; right column) within the simulation region. Each row shows a different simulation and the color corresponds to three different points in time (SFE $=0, 5, 10 \%$). As in Fig.~\ref{fig:rate_dens_all}, the transition density ($s_{\mathrm{t}}$) from \cite{Appel+2022} is overplotted.  The sink formation density threshold ($s_{\mathrm{sink}}$) for the simulation is also shown. 
\label{fig:pdf_sum}}
\end{figure}

Similar to the upper panels of Fig.~\ref{fig:rate_dens_all}, Fig.~\ref{fig:pdf_sum} shows a comparison between the overall PDF and the expanding or compressing PDF for all four physics cases (we do not include run \textsc{GTMJ} here). 
We show three different points in time for each simulation, corresponding to just before the formation of the first sink particle (SFE $= 0 \%$), the approximate mid-point of each simulation (SFE = $5 \%$), and the end of each simulation (SFE $= 10\%$). 
For all four physics cases and all three points in time, the same trend is apparent. 
At high densities, the overall density PDF is well matched by the compressing gas PDF but is much higher than the expanding gas PDF. 
However, at low densities, the overall PDF is well matched by the expanding gas PDF but diverges from the compressing gas PDF. 
This confirms that most of the expanding gas is at low densities while the compressing gas is predominantly at high densities. 
The transition between these regimes is continuous and fairly gradual.

\bibliographystyle{aasjournal}
\bibliography{bibliography.bib,vs.bib}{}

\end{document}

%% file: citation_fix.tex
\usepackage{etoolbox}
\makeatletter
\patchcmd{\NAT@citex}
  {\@citea\NAT@hyper@{\NAT@nmfmt{\NAT@nm}\NAT@date}}
  {\@citea\NAT@nmfmt{\NAT@nm}\NAT@hyper@{\NAT@date}}
  {}
  {}
\patchcmd{\NAT@citex}
  {\@citea\NAT@hyper@{%
     \NAT@nmfmt{\NAT@nm}%
     \hyper@natlinkbreak{\NAT@aysep\NAT@spacechar}{\@citeb\@extra@b@citeb}%
     \NAT@date}}
  {\@citea\NAT@nmfmt{\NAT@nm}%
   \NAT@aysep\NAT@spacechar%
   \NAT@hyper@{\NAT@date}}
  {}
  {}
\patchcmd{\NAT@citex}
  {\@citea\NAT@hyper@{%
     \NAT@nmfmt{\NAT@nm}%
     \hyper@natlinkbreak{\NAT@spacechar\NAT@@open\if*#1*\else#1\NAT@spacechar\fi}%
       {\@citeb\@extra@b@citeb}%
     \NAT@date}}
  {\@citea\NAT@nmfmt{\NAT@nm}%
   \NAT@spacechar\NAT@@open\if*#1*\else#1\NAT@spacechar\fi%
   \NAT@hyper@{\NAT@date}}
  {}
  {}
\makeatother